\documentclass{amsart}
\usepackage{amssymb}
\usepackage{fullpage}
\usepackage{graphicx}
\input pictex.sty
\newtheorem{thm}{Theorem}[section]

\newtheorem{lem}[thm]{Lemma}
\newtheorem{prop}[thm]{Proposition}
\theoremstyle{definition}

\theoremstyle{remark}

\numberwithin{equation}{section}

\newcommand{\cb}{\mathbb{C}}

\newcommand{\lam}{\lambda}

\newcommand{\al}{\alpha}

\begin{document}

\title{Random Matrices, Graphical Enumeration and the Continuum Limit of Toda Lattices}%
\author{N. M. Ercolani}%
\address{Dept. of Math, Univ. of Arizona, 520-621-2713, FAX: 520-626-5186}%
\email{ercolani@math.arizona.edu}%
\author{K. D. T-R McLaughlin}%
\address{Dept. of Math., Univ. of Arizona}%
\email{mcl@math.arizona.edu}
\author{V. U. Pierce}%
\address{Dept. of Math., The Ohio State University}%
\email{vpierce@math.ohio-state.edu}

\thanks{K. D. T-R McLaughlin was supported in part by NSF
grants DMS-0451495 and DMS-0200749, as well as a NATO
Collaborative Linkage Grant "Orthogonal Polynomials: Theory,
Applications, and Generalizations" Ref no. PST.CLG.979738. N. M.
Ercolani and V. U. Pierce were supported in part by NSF grants
DMS-0073087 and DMS-0412310.}


\begin{abstract}
In this paper we derive a hierarchy of differential equations
which uniquely determine the coefficients in the asymptotic
expansion, for large $N$, of the logarithm of the partition
function of $N \times N$ Hermitian random matrices. These
coefficients are generating functions for graphical enumeration on
Riemann surfaces. The case that we particularly consider is for an
underlying measure that differs from the Gaussian weight by a
single monomial term of degree $2\nu$. The coupling parameter for
this term plays the role of the independent dynamical variable in
the differential equations. From these equations one may deduce
functional analytic characterizations of the coefficients in the
asymptotic expansion. Moreover, this ode system can be solved
recursively to explicitly construct these coefficients as
functions of the coupling parameter.  This analysis of the fine
structure of the asymptotic coefficients can be extended to
multiple coupling parameters and we present a limited illustration
of this for the case of two parameters.
\end{abstract}
\maketitle

\section{Motivation and Background}

The study of the Unitary Ensembles (UE) of random matrices
\cite{Mehta}, begins with a family of probability measures on the
space of $N \times N$ Hermitian matrices. The measures are of the
form

$$
d\mu_\textbf{t} = \frac{1}{\widetilde{Z}_N}\exp\left\{-N \mbox{ Tr
} [V_\textbf{t}(M)]\right\} dM,
$$
where the function $V_\textbf{t}$ is a scalar function, referred
to as the potential of the external field, or simply the "external
field" for short.  Typically it is taken to be a polynomial, and
written as follows:

$$
V_\textbf{t} = \frac{1}{2} \lambda^{2} + \sum_{j=1}^{\upsilon}
t_{j}\lambda^{j}.
$$

The partition function $\widetilde{Z}_N$, which appears as a
normalization factor in the UE measures, plays a central role in
random matrix theory and its applications. It can be reduced to an
integration over the eigenvalues which takes a form proportional
to the integral (\ref{I.001}), below, for the particular case when
$k=N$.

When all the coefficients $t_k$ in the external field are set
equal to zero the associated ensemble, corresponding to $\mu_0$,
is called the \emph{Gaussian Unitary Ensemble} (GUE). Many
simplifications occur in the Gaussian case (see \cite{EM} for
explanations of any unfamiliar terms):

\begin{enumerate}
\item  The partition function, when all $t_k$ vanish, is a
Gaussian integral, and can be evaluated exactly.

 \item The matrix moments, $\int \{\mbox{ Tr }(M^j)\}^k d\mu_0 (M)$, can be evaluated, using
 Wick's lemma, in terms of pair correlations of the matrix
entries of $M$ which are complex normal random variables.

\item The terms in these Wick coupling expansions are, in the
manner of Feynman diagrams, in 1-1 correspondence with certain
labelled, oriented graphs.
\end{enumerate}

These observations led to the conjecture \cite{BIZ, Difrancesco95}
that the logarithm of the  partition function has an asymptotic
expansion of the form:

$$
\log
\left(\frac{\widetilde{Z}_{N}(\bf{t})}{\widetilde{Z}_{N}(\bf{0})}
\right) = N^{2} e_{0}({\bf{t}}) + e_{1}({\bf{t}}) +
\frac{1}{N^{2}} e_{2}({\bf{t}}) + \cdots
$$
where the coefficients $e_g(\textbf{t})$ should be locally
analytic functions of {\bf{t}}. The Taylor coefficients of $e_g$
should enumerate topologically distinct labelled, connected
oriented graphs that can be embedded into a Riemann surface of
genus g in such a way that the complement of the graph in the
surface is a disjoint union of contractible cells. Such a
construction is referred to as a \emph{g-map} (see section
\ref{g-maps} for a precise definition). The $e_g(t)$ are
generating functions for counting the number of \emph{g}-maps with
given numbers of vertices of specified valence. This conjecture
was proven in \cite{EM} for appropriate domains (see below). The
present paper builds on these results to present a more detailed
description of the coefficients $e_g(\textbf{t})$ and related
generating functions.
\medskip

More precisely, our interest is to develop a systematic, rigorous
description of the fine structure for the large $N$ asymptotics of
the following family of integrals:

\begin{eqnarray}
\label{I.001}
& & Z^{(k)}_{N}(t_{1},t_{2}, \ldots, t_{\upsilon}) =\\
\nonumber & & \int \cdots \int \exp{ \left\{
-N^{2}\left[\frac{1}{N} \sum_{j=1}^{k} V(\lambda_{j}; \ t_{1},
\ldots, t_{\upsilon})  - \frac{1}{N^{2}} \sum_{j\neq \ell} \log{|
\lambda_{j} -
\lambda_{\ell} | } \right] \right\} } d^{k} \lambda, \\
& & V(\lambda; \ t_{1}, \ldots, t_{\upsilon} ) = V_{{\bf
t}}(\lambda) = V(\lambda) = \frac{1}{2} \lambda^{2} +
\sum_{j=1}^{\upsilon} t_{j}\lambda^{j}.
\end{eqnarray}
where the parameters $\{t_{1},\ldots, t_{\upsilon}\}$ are assumed
to be such that the integral converges.  For example, one may
suppose that $\upsilon$ is even, and $t_{\upsilon}>0$.  We will
sometimes refer to the following set of ${\bf{t}} = (t_{1},
\ldots, t_{\upsilon})$ for which (\ref{I.001}) converges. For any
given $T>0$ and $\gamma
> 0$, define
\begin{equation*}
\mathbb{T}(T,\gamma) = \{ {\bf{t}} \in \mathbb{R}^{\upsilon}: \
|{\bf{t}} | \le T, \ t_{\upsilon} > \gamma \sum_{j=1}^{\upsilon-1}
|t_{j}|\}.
\end{equation*}
The parameter $k$ is an integer that grows with $N$ in such a way
that $\lim_{N\to\infty} k/N = x$, where $x$ is a finite non-zero
value whose role will be specified more precisely later.
\medskip

In this paper we derive a hierarchy of differential equations
which uniquely determine the coefficients in the asymptotic
expansion of $\log Z_N^{(N)}$ for monic even coupling parameters;
i.e., we present the $e_g(t_{2\nu})$, for arbitrary $\nu$, as
solutions to a system of ordinary differential equations. From
this one can deduce functional analytic characterizations of these
coefficients. Moreover, this ode system can be solved recursively
in \emph{g} to explicitly construct $e_g(t_{2\nu})$. We illustrate
this process by constructing closed form expressions for
$e_g(t_{2\nu})$, in which $\nu$ appears as a parameter, for low
values of \emph{g}. This analysis of the fine structure of the
$e_g$ can be extended to multiple coupling parameters and we
present a limited illustration of this for the case of two
parameters: $e_g(t_{2\nu_1}, t_{2\nu_2})$.
\bigskip

{\bf Remark} In \cite{Ambjorn}, the so-called "Loop Equation"
method is used to obtain some information about the fine structure
of the coefficients.  This approach is based on a formal
derivation of a hierarchy of equations for the Cauchy transform of
the mean density of eigenvalues.  This interesting approach is
unsatisfactory in that it relies on several interchanges of
singular limits whose justification requires analytical
considerations beyond the existence of the complete asymptotic
expansion of the partition function.  These analytical
considerations are the subject of a forthcoming paper by Ercolani
and McLaughlin \cite{ErMcLNew}. $\Box$

\subsection{Leading Order Asymptotics}

The leading order behavior of $Z^{(k)}_{N}(t_{1},t_{2}, \ldots,
t_{\upsilon})$ is rather classical, and is known for a very wide
class of external fields $V$ (see, for example, \cite{Johansson}).
We will require the following result.

\begin{thm}\label{thm:lead}
There is $T_{0}>0$ and $\gamma_{0}>0$ so that for all $\bf{t} \in
\mathbb{T}(T,\gamma)$, $x\in [1/2,1]$, and $k/N \rightarrow x$ as
$k,N \rightarrow \infty$, the following holds true:
\begin{enumerate}
\item
\begin{eqnarray}
\label{eq:lead01} && \lim_{N \to \infty} \frac{1}{k^{2}}
\log\{Z^{(k)}_{N}(t_{1},t_{2}, \ldots, t_{\upsilon}) \} = -I (x,
t_{1}, \ldots, t_{\upsilon})
\end{eqnarray}
where
\begin{eqnarray}
\label{eq:lead02} I (x, t_{1}, \ldots, t_{\upsilon}) = &&
\inf_{\text{Borel measures  }\mu ,\mu\geq 0, \int d\mu =1} \left[
\frac{1}{x} \int V(\lam ) d\mu (\lam ) -\int\int \log | \lam -\mu
|\, d\mu (\lam )\, d\mu (\eta )\right] .
\end{eqnarray}
\item There is a unique measure $\mu_V$ which achieves the infimum
defined on the right hand side of (\ref{eq:lead02}). This measure
is absolutely continuous with respect to Lebesgue measure, and
  \begin{align*}
d\mu_{V} &= \psi \, d\lam ,\\
\psi (\lam ) &= \frac{1}{2 \pi} \chi_{(\al ,\beta )} (\lam )
\sqrt{(\lam -\al )(\beta -\lam )}\, h(\lam ),
  \end{align*}
where $h(\lam )$ is a polynomial of degree $\upsilon -2$, which is
strictly positive on the interval $[\alpha,\beta]$ (recall that
the external field $V$ is a polynomial of degree $\upsilon$). The
polynomial $h$ is defined by
  \[
h(z) = \frac{1}{2\pi i x}\, \oint\, \frac{V'(s)}{\sqrt{(s-\al )}
\sqrt{(s-\beta )}} \, \frac{ds}{s-z}
  \]
where the integral is taken on a circle containing $(\al ,\beta )$
and $z$ in the interior, oriented counter-clockwise.

\item There exists a constant $l$, depending on $V$ such that the
following variational equations are satisfied by $\mu_V$:
\begin{eqnarray}
\nonumber  \int 2 \log|\lambda - \eta|^{-1} d\mu_V(\eta) + x^{-1}V(\lambda)
&\geq& l \,\,\, \mbox{for}\,\,\, \lambda \in \mathbf{R}\backslash\mbox{supp}(\mu_V) \\
  \int 2 \log|\lambda - \eta|^{-1} d\mu_V(\eta) + x^{-1}V(\lambda)
  &=& l \,\,\, \mbox{for}\,\,\, \lambda \in \mbox{supp}(\mu_V). \label{Var}
\end{eqnarray}

\item The endpoints $\al$ and $\beta$ are determined by the
equations
  \begin{align*}
\int^{\beta}_{\al}\, \frac{V'(s)}{\sqrt{(s-\al )(\beta -s)}}\, ds &= 0\\
 \int^{\beta}_{\al}\, \frac{sV'(s)}{ \sqrt{(s-\al )(\beta -s)} }\, ds &=
2\pi x.
  \end{align*}

\item The endpoints $\al (x,{\bf t}\, )$ and $\beta (x,{\bf t}\,
)$ are actually analytic functions of ${\bf t}$ and $x$, which
possess smooth extensions to the closure of $\{ x,{\bf t}: x \in
[1/2,1], {\bf t} \in \mathbb{T}(T,\gamma) \}$.  They also satisfy
$-\alpha(1,{\bf 0}) = \beta(1,{\bf 0}) = 2$. In addition, the
coefficients of the polynomial $h(\lam )$ are also analytic
functions of ${\bf t}$ and $x$, with smooth extensions to the
closure of $\{ x,{\bf t}: x \in [1/2,1], {\bf t} \in
\mathbb{T}(T,\gamma) \}$, with
  \[
h(\lam ,x=1, {\bf t}={\bf 0}) = 1.
  \]
\end{enumerate}
\end{thm}
\bigskip

{\bf Remark}  The variational problem appearing in
(\ref{eq:lead02}) is a fundamental component in the theory of
random matrices, as well as integrable systems and approximation
theory.  It is well known, (see, for example, \cite{SaffTotik}),
that under general assumptions on $V$, the infimum is achieved at
a unique measure $\mu_{V}$, called the equilibrium measure. For
external fields $V$ that are analytic in a neighborhood of the
real axis, and with sufficient growth at $\infty$, the equilibrium
measure is supported on finitely many intervals, with density that
is analytic on the interior of each interval, behaving at worst
like a square root at each endpoint, (see \cite{DKM} and
\cite{DKMVZ3}). $\Box$
\medskip

{\bf Remark}  We call the reader's attention to the parameter,
$x$, in the formulation of the variational problem.  We will
consider the variational problem for $x \in (0,1]$, and we are
particularly interested in $x$ near $1$.  This parameter
represents the asymptotic ratio of $k$ to $N$:  $x =  \lim_{N \to
\infty} k/N$. $\Box$
\medskip

{\bf Remark}  For a proof of (\ref{eq:lead01}), we refer the
reader to \cite{Johansson}, however this result is commonly known
in the approximation theory literature. $\Box$
\medskip

{\bf Remark} It will prove useful to adapt the following
alternative presentation for the function $\psi$:
\begin{eqnarray} \label{psi}
\psi(\lambda) = \frac{1}{2 \pi i} R_{+}(\lambda) h(\lambda), \
\lambda \in (\alpha, \beta),
\end{eqnarray}
where the function $R(\lambda)$ is defined via $R(\lambda)^{2} = (
\lambda - \alpha) ( \lambda - \beta)$, with $R(\lambda)$ analytic
in $\cb \setminus [\alpha, \beta]$, and normalized so that $R(
\lambda) \sim \lambda$ as $\lambda \to \infty$.  The subscript
$\pm$ in $R_{\pm}(\lambda)$ denotes the boundary value obtained
from the upper (lower) half plane. $\Box$

\subsection{Complete Asymptotic Expansion}

\vskip 0.2in

In \cite{EM} it was established that a complete large $N$
asymptotic expansion of \ref{I.001} exists.  In this paper we will use
a straightforward generalization of this result:

\begin{thm}\label{EQMSTHM}
There is $T>0$ and $\gamma > 0$ so that for $\bf{t} \in
\mathbb{T}(T,\gamma)$, and $x = k/N$ in a neighborhood of $x=1$, one has the
$N \to \infty$ asymptotic
expansion
\begin{eqnarray}
\label{I.002} \ \ \ \log{
\left(\frac{Z^{(k)}_{N}(\bf{t})}{Z^{(k)}_{N}(\bf{0})} \right)} =
k^{2} e_{0}(x, {\bf{t}}) + e_{1}(x, {\bf{t}}) + \frac{1}{k^{2}}
e_{2}(x, {\bf{t}}) + \cdots.
\end{eqnarray}
The meaning of this expansion is:  if you keep terms up to order
$k^{-2h}$, the error term is bounded by $C k^{-2h-2}$, where the
constant $C$ is independent of $x$ and $\bf{t}$ for all $\bf{t}
\in \mathbb{T}(T,\gamma)$ and for all $x$ in the neighborhood of
1.  For each $j$, the function $e_{j}(x, {\bf{t}})$ is an analytic
function of the (complex) vector $(x, {\bf{t}})$, in a
neighborhood of $(1, {\bf{0}})$. Moreover, the asymptotic
expansion of derivatives of $\log{ \left( Z^{(k)}_{N} \right)}$
may be calculated via term-by-term differentiation of the above
series. $\Box$
\end{thm}
\bigskip

{\bf Remark} In \cite{EM}, this result was established in the case
where $x=1$, and under the assumption that ${\bf t} \in
\mathbb{T}(T, \gamma)$, for $T$ small enough, and $\gamma$ large
enough, so that Theorem \ref{EQMSTHM} holds true. Under these
assumptions, Theorem \ref{thm:01} (below) was established.
However, as observed in \cite{EM} (Remark 2.1, page 2), the domain
so defined is by no means the largest domain where the asymptotic
expansion can be rigorously established.  All that is required is
the existence of a path through the space of parameters (values of
$x$ and  ${\bf t}$) connecting  $(x, {\bf t})$ to $(1,{\bf 0})$ in
such a way that all along the path, the associated equilibrium
measure is supported on a single interval, with strict variational
inequality on the support, strict positivity on the interval of
support, and vanishing like a square root at both endpoints of the
support.  The collection of all such values of $(x,{\bf t})$
defines a suitable candidate for a maximal domain, and the proof
contained in \cite{EM} can easily be extended to show that the
asymptotic expansion of the partition function holds on the
interior of such a domain.  In particular, the above Theorem may
be easily deduced along these lines. $\Box$
\medskip

{\bf Remark} Recently, Bleher and Its \cite{BI} have carried out a
similar asymptotic expansion of the partition function for a
1-parameter family of external fields.  A very interesting aspect
of their work is that they establish the nature of the asymptotic
expansion of the partition through a critical phase transition.
$\Box$

\subsection{Graphical Enumeration and the Partition Function
Expansion}\label{g-maps}

Our goal in the work we present here is to establish
\emph{analytical} characterizations of the coefficients $e_g$ and,
when possible, to derive explicit expressions for these
coefficients. This is what we mean by the \emph{fine structure} of
the expansion.

In addition to providing the first proof of the asymptotic
expansion described in Theorem \ref{EQMSTHM}, \cite{EM} also
provides a very detailed explanation of the connection between the
asymptotic expansion and \emph{enumerative geometry}, originally
investigated by physicists in the 70s and 80s (see, for example,
\cite{BIZ}, \cite{Difrancesco95}, and references contained
therein).  Equipped with the existence of the asymptotic expansion
(and the subsequent result that it may be differentiated term by
term), one shows that there is a \emph{geometric} characterization
of each $e_g$ as a generating function for enumerating
topologically distinct embeddings of graphs into Riemann surfaces
of genus $g$.

A {\it map} $D$ on a compact, oriented connected surface X is a
pair $D = (K(D), [\imath])$ where
\begin{enumerate}
\item $K(D)$ is a connected 1-complex; \item $[\imath]$ is an
isotopical class of inclusions $\imath:K(D) \rightarrow X$; \item
the complement of $K(D)$ in $X$ is a disjoint union of open cells
(faces); \item the complement of $K_0(D)$ (vertices) in $K(D)$ is
a disjoint union of open segments (edges).
\end{enumerate}
The $e_g$ enumerate labelled maps. To be precise we
introduce the notion of a \emph{g-map} which is a map in which the
surface X is the closed, oriented Riemann surface of genus g and
which in addition carries a labelling (ordering) of the vertices.

\begin{thm} \label{thm:01}\cite{EM}
\label{GeomAsymp} The coefficients in the asymptotic expansion
(\ref{I.002}) satisfy the following relations.  Let $g$ be a
nonnegative integer.  Then
\begin{equation} \label{genusexpA}
e_g(t_1 \dots t_\upsilon) =  \sum_{n_j\geq 1}\frac{1}{n_1!\dots
n_\upsilon!} (-t_1)^{n_1}\dots
(-t_\upsilon)^{n_\upsilon}\kappa_g(n_1,\dots,n_\upsilon)
\end{equation}
in which each of the coefficients $\kappa_g(n_1, \dots,
n_\upsilon)$ is the number of g-maps with $n_j$  j-valent vertices
for $j=1,\dots,\upsilon$.
\end{thm}

\subsection{Outline}

The organization of this paper is as follows: In section 2 we
present the new results concerning the fine structure of the $e_g$
and related generating functions that will be proven and further
explained in the remainder of the paper.

Section 3 is concerned with the leading order term, $e_0$. The
results here are fundamental for the characterization of all the
higher order terms. We derive closed form expressions for $e_0$ as
a function of each of the valence coupling parameters $t_{2\nu}$.
We also relate these evaluations directly and explicitly to the
enumeration of planar graphs.

In section 4 a continuum limit of the Toda Lattice hierarchy is
rigorously derived in which the hierarchy of Toda \emph{times}
corresponds to the valence coupling parameters $t_{2\nu}$. This
continuum limit is then used to derive another hierarchy of
differential equations whose solutions are the $e_g$.

Finally in section 5 we show how the differential equations
derived in the previous section are used to inductively generate
explicit expressions for the $e_g$. From this we characterize the
function-theoretic structure of the $e_g$ as well as present
explicit formulae for the $e_g$ for low values of $g$. We also
show how our results may be extended to the case of multiple
times.

\section{Results}

\noindent For $e_0$ we have explicit formulas for monic even times

\begin{thm}\label{II.001}
For potentials $V$ of the form $V = \frac{1}{2}\lambda^2 +
t_{2\nu}\lambda^{2\nu}$, the asymptotic expansion \eqref{I.002}
holds true for all $t_{2 \nu}\ge0$, and in addition, we have the
explicit formula
$$e_0 = \eta(z-1)(z-r) + \frac{1}{2} \log(z)$$
where
  \begin{align*}
\eta &= \frac{(\nu-1)^2}{4\nu(\nu+1)},\\
r &= \frac{3(\nu+1)}{\nu-1},\\
%
%
z &= \frac{\beta^2}{4}.
  \end{align*}
Here $4z$ can be interpreted as the global analytic continuation
of $\beta^2$ which determines the support $(-\beta, \beta)$ of the
equilibrium measure. The variable $z$ is locally an analytic
function of $t_{2\nu}$, which satisfies the algebraic relation
\begin{eqnarray}
\nonumber
1 = z + 2 \nu \binom{2\nu-1}{\nu-1} x^{\nu-1} t_{2 \nu} z^\nu.
\end{eqnarray}
The singularities of $e_0$ occur at $z = 0$ and $z = \infty$. The
time derivative
\begin{eqnarray*}
  \frac{\partial e_0}{\partial t_{2\nu}} = \left(\begin{array}{c}
  2\nu-1\\
  \nu-1\\
\end{array}\right) z^\nu\left((\nu-1)z - (\nu+1)\right)
\end{eqnarray*}
is polynomial in $z$.

One also has a local analytical representation (here the index
$n_{2\nu}$ is replaced by $n$, so that
$\kappa_{0}(0,\ldots,0,n_{2\nu})$ becomes $\kappa_{0}(n)$),
\begin{align*}
e_0(t_{2\nu})    &= \sum_{j=1}^\infty \kappa_{0}(n)
\frac{(-t_{2\nu})^n}{n!},\\
\kappa_{0}(n) &= (c_\nu)^n \frac{(\nu n -1)!}{((\nu-1)n+2)!},\\
c_\nu &= 2\nu \left(\begin{array}{c}
  2\nu-1\\
  \nu-1\\
\end{array}\right),
  \end{align*}
where $\kappa_{0}(n)=\kappa_{0}(n_{2\nu})$ is the generating
function for $2\nu$-valent 0-maps.
\end{thm}

To get a handle on how the higher coefficients $e_g$ depend on the
parameters $t = t_{2\nu}$ we exploit a remarkable relation between
the partition function $Z^{(N)}_{N}(\bf{t})$ and the solutions to
the hierarchy of completely integrable semi-infinite Toda lattice
equations. These differential equations may be succinctly
expressed through the semi-infinite tri-diagonal matrix
\begin{equation}\label{Jac}
\mathcal{L} = \left(%
\begin{array}{ccccc}
  0 & 1 & 0 & 0 & \cdots \\
  b_0^2 & 0 & 1 & 0 & \cdots \\
  0 & b_1^2 & \ddots & \ddots & \ddots \\
  0& 0 & \ddots& 0 & 1 \\
  \vdots & \vdots & \ddots & b_n^2 & \ddots \\
\end{array}%
\right).
\end{equation}
The Toda Lattice system at level $2\nu$ can then be defined as
\begin{eqnarray}\label{beqns1}
\frac{1}{2} \frac{db_k^2}{d\xi} &=& (\mathcal{L}^{2\nu})_{k+2,k} -
(\mathcal{L}^{2\nu})_{k+1,k-1},\\
\label{beqns1.1}(\mathcal{L}^{2\nu})_{k+1,k-1} &=& \sum_{i_1,i_2, \ldots,
i_{2\nu+1};
 |i_{j+1}-i_j|=1; i_1=k+1, i_{2\nu+1}=k-1} \mathcal{L}_{k+1,i_2}\mathcal{L}_{i_2,i_3}\ldots \mathcal{L}_{i_{2\nu},k-1}.
\end{eqnarray}
The sum here is indexed by walks of length $2\nu$ along the 1D
integer lattice from  $k+1$ to $k-1$. The solution of this system
may be expressed directly in terms of the partition function
$Z_k(t_1, t) = Z_k^{(k)}$ associated to the potential $V =
\frac{1}{2}\lambda^2 + t_1 \lambda + t\lambda^{2\nu}$:

\begin{equation}\label{TodaSoln}
b_k^2(\xi) = k (\frac{1}{2k^2})\frac{d^2}{d{t_1}^2}\log
Z_k({t_1},s)_{{t_1}=-k^{-1/2}\xi_1=0, s=2\xi k^{\nu-1}}.
\end{equation}
As a dynamical system, one is really considering an initial value
problem, with
\begin{eqnarray}
b_{k}(0)^{2} = k.
\end{eqnarray}
 We can now state our next main result which characterizes the
continuum limit of the Toda lattice hierarchy.

\begin{thm}\label{II.002}
For all $t\ge0$, $b_k^2$ has a valid
asymptotic expansion of the form
$$b_k^2 \simeq k(z_0(s)+\frac{1}{k^2}z_1(s)+\frac{1}{k^4}z_2(s)+\cdots)$$
where $s=-2k^{\nu-1}t$. The terms of this expansion are determined
by the following partial differential scheme:
  \begin{align*}
f_s & \left.=  c_\nu f^\nu f_w +
\frac{1}{k^2}F_1^{(\nu)}(f,f_w,f_{ww},f_{www})+\cdots +
\frac{1}{k^{2g}} F_g^{(\nu)}(f, f_w, f_{w^{(2)}}, \cdots,
f_{w^{(2g+1)}}) + \cdots \ \right|_{\textrm{evaluated at} \,\,  w=1};\\
\textrm{where}& \\
f(s,w) &= f_0(s,w) + \frac{1}{k^2}f_1(s,w)+ \cdots + \frac{1}{k^{2g}}f_g(s,w) + \cdots,\,\,\, \textrm{and}\\
f(s,1) &= z_0(s)+\frac{1}{k^2}z_1(s)+\frac{1}{k^4}z_2(s)+\cdots,\\
f_g(s,w)&= w^{1-2g}z_g(w^{\nu-1}s).
  \end{align*}
Note that $b_{k}^{2}$ and $k f(s,1)$ possess the same asymptotic expansion.

\medskip
\noindent  The forcing term
$F_j^{(\nu)}(\cdots)|_{w=1}$ is a homogeneous multi-nomial of
degree $\nu + 1$ in the $f_{w^{(r)}}$ which does not contain any
instances of $z_\alpha$ for $\alpha \geq j$.

These forcing terms have the following form:

$$F_g^{(\nu)} = \sum_{V:|V|= 2g+1 \ni \; \rho(V) \leq \nu+1}d_V^{(\nu,g)}f^{\nu -
\rho(V)+1}\prod_{j=1}^{2g+1} \left( \frac{f_{w^{(j)}}}{j!}
\right)^{r_j(V)}$$ where $V=\bigcup_{m=1}^{\rho(V)}V_m$ is a
partition of $2g+1$; $r_j$ is the number of times a "part", $V_m$,
of cardinality $|V_m|=j$ appears in the partition; $\rho = \rho(V)
= \sum r_j(V)$; and
\begin{align*}
d_V^{(\nu,g)} &= \frac{1}{\prod_{j=1}^{2g+1}r_j!}\sum_{1\leq i_1<
\cdots< i_{\rho(V)}\leq 2\nu}\left( \textrm{coeff of}\,\,
x^{\nu-\rho(V)+1} \textrm{in}\,\, P(x) - \textrm{coeff of}\,\,
x^{\nu-\rho(V)+1}
\textrm{in}\,\, Q(x)\right); \textrm{where},\\
P(x) &= \sum_{\sigma\in \mathcal{S}}\prod_{m=1}^{\rho(V)}
\left(i_m-2\sum_{s=1}^m(1+x_s\frac{\partial}{\partial x_s})+ 2
\right)^{|V_{\sigma(m)}|}\\
& \cdot (1+x_1)^{i_1-1}\cdots (1+x_s)^{i_s-i_{s-1}-1}\cdots
(1+x_\rho)^{i_\rho-i_{\rho-1}-1}(1+x_{\rho+1})^{2\nu-i_\rho}|_{x_\mu=x},\\
Q(x) &= \sum_{\sigma\in \mathcal{S}}\prod_{m=1}^{\rho(V)}
\left(i_m-2\sum_{s=1}^m(1+x_s\frac{\partial}{\partial
x_s})+1\right)^{|V_{\sigma(m)}|}\\
 & \cdot (1+x_1)^{i_1-1}\cdots
(1+x_s)^{i_s-i_{s-1}-1}\cdots
(1+x_{\rho+1})^{2\nu-i_\rho}|_{x_\mu=x}.
\end{align*}
  \end{thm}
\bigskip

{\bf Remark}
We refer to the above as a \emph{partial differential
scheme} because it signifies not an equation to be solved but
rather a prescription for generating a hierarchy of ordinary
differential equations for the $z_g$. The ode hierarchy is
constructed from the scheme as follows. The ode at level $g$ is
obtained by replacing the expansion $f$ and its $w$-derivatives,
$f_w,f_{ww}$, etc., in the pde scheme by their order $g$
truncations. Then, equating the coefficients of all terms of order
$k^{-2g}$ in this truncated scheme and setting $w=1$ yields a
$k$-independent ode in $s$ which is the $g^{th}$ equation of the
continuum-Toda hierarchy. This is an ode for $z_g(s)$ in terms of
$z_j(s)$ for $j \leq g$.
$\Box$

The first equation in the hierarchy (i.e., the one coming from the
$k^0$-coefficients of the above scheme) is a nonlinear ODE for
$z_0$:
\begin{equation*}
z_0'(s) = c_\nu z_0(s)^\nu \left( z_0(s) + (\nu-1) s
z_0'(s)\right),
\end{equation*}
with the initial condition $z_0(0) = 1$. This ODE is solved
implicitly by a solution to the algebraic equation
\begin{equation} \label{II.005}
1 = z_0(s) - c_\nu s z_0(s)^\nu .
\end{equation}
As indicated in Theorem \ref{thm:lead} (5), this relation can also be derived
directly from the characterization of the equilibrium measure.
Relation (\ref{II.005}) allows us to write $s$ as well as
derivatives of $z_0$ as rational functions of $z_0$. This will be
exploited to arrive at the explicit representations of the $z_g$
given below.

The $k^{-2g}$ equation in the hierarchy is linear in $z_g$, and
can be written as:
\begin{equation} \label{II.006}
z_g'(s) = c_\nu \left( f_0^\nu {f_g}_w + \nu f_0^{\nu-1} f_g
{f_0}_w \right)_{w=1} + \mbox{Forcing}_g|_{w=1}
\end{equation}
where
\begin{equation}\label{Forcing}
\mbox{Forcing}_g = \left(\frac{c_\nu}{\nu+1}
\frac{\partial}{\partial w} \sum_{\begin{array}{c}
  0 \leq i_j < g \\
  i_1 + \dots + i_{\nu+1} = g\\
\end{array}} f_{i_1}\cdots f_{i_{\nu +1}}\right) + F_1^{(\nu)}[2g-2] +
F_2^{(\nu)}[2g-4] + \cdots + F_{g}^{(\nu)}[0],
\end{equation}
and  $F_\ell^{(\nu)}[2r]$ denotes the coefficient of $k^{-2r}$ in
$F_\ell^{(\nu)}$. We note that the terms in
$\mbox{Forcing}_g|_{w=1}$ depend only on $z_j, j<g$ and their
derivatives.
\medskip

{\bf Remark} Amongst the results on the above singular limit of
the Toda lattice in the literature, we remark that the recent work
of Bloch, Uribe, and Golse \cite{BGU} is related, in the sense
that in their work, through the use of the theory of Toeplitz
operators,  the authors establish the existence of an asymptotic
expansion for a continuum limit of a finite dimensional Toda
lattice. $\Box$
\medskip

For the case of planar maps ($g=0$) such generating functions have
received significant attention recently \cite{BDG, B-MS}. Our
explicit calculations presented later can provide a basis for
extending these studies.
\medskip

Before proceeding to the statement of our next result we need to
introduce a scale of function classes that will enable us to
describe the functional nature of the coefficients $z_g$ that we
have just introduced as well as that of the generating functions
$e_g$. We will refer to the classes as \emph{iterated integrals of
rational functions} or \emph{iir} for short. These classes are
defined inductively in terms of the variable $z = z_0$ regarded as
an independent variable as follows. To begin with, the class
contains rational functions of $z$. One then adds integrals of
these rational functions with respect to $dz$. Next one considers
the vector space of polynomials in products of these integrals
over the field of rational functions in $z$ and augments the space
by integrals, with respect to $dz$ of these functions. Then take
the vector space of polynomials in these latter integrals and add
integrals of these. One continues this iterative process up to any
given finite stage. These are the classes of functions we refer to
as \emph{iir}. In our case the rational functions at any stage
will be restricted to the sub-ring of functions whose poles are
located at either $z=0$ or $z = \nu/(\nu-1)$ for a fixed value of
$\nu$. These classes of functions will certainly include the class
generated by \emph{polylogarithms} \cite{Erdelyi} but may be
larger.


\begin{thm}\label{II.003thm}

\begin{enumerate}

\item The coefficient $z_{g}$ is of class iir in $z_{0}$
with singularities only possible at $z_0 = 0$ and
$z_0=\nu/(\nu-1)$.

\item
The coefficient $z_{g}$ is more explicitly presented as a function of
$z_{0}$ through the following integral solution of Equation
(\ref{II.006}):
\begin{equation*}
z_g(s) = \frac{z_0(s)^{2(1-g)}}{\nu-(\nu-1)z_0(s)} \int_1^{z_0(s)}
\frac{(\nu-(\nu-1)y)}{c_\nu y^{\nu+3-2g} }
\mbox{\textup{Forcing}}_g(y) dy.
\end{equation*}
\item In the above equation, $\mbox{\textup{Forcing}}_g$, formerly a
function of a great many arguments, is in fact a function of $z_0(s)$ alone,
which will henceforth be denoted as $\mbox{\textup{Forcing}}_g(z_0)$.
\end{enumerate}
\end{thm}

We also derive a hierarchy of differential equations for the
$e_g(t_{2\nu})$ with data given in terms of the $z_j's$.
\begin{thm}\label{II.003Bthm}
The $g$'th equation in the hierarchy of equations governing
$e_g(t_{2\nu})$ is \begin{eqnarray}\label{II.007}
\frac{\partial^2}{\partial w^2} \left[ w^{2-2g}
e_g\left(-w^{\nu-1} s\right)
  \right]\bigg|_{w=1} &=& -\sum_{n
=1}^{g} \frac{2}{(2n+2)!} \frac{\partial^{(2n+2)}}{\partial
  w^{(2n+2)}}
\left[
  w^{2-2(g-n)} e_{g-n}(-w^{\nu-1} s)  \right] \bigg|_{w=1}\\
  \nonumber  &+& \mbox{the}\;
  k^{-2g} \; \mbox{term of} \; \log\left( \sum_{n = 0}^\infty \frac{1}{k^{2n}}
  z_n(s) \right).
\end{eqnarray}
\end{thm}
Equation (\ref{II.007}) determines $e_g(-s)$ from a second order
differential equation for $e_g$ with forcing terms depending on
$e_n,\,\, n<g, \;$ $z_n, \,\, n\leq g$, and their derivatives.
\bigskip

{\bf Remark} Observe that the RHS of (\ref{II.007}) is a function
of $z_0(s)$ which will henceforth be denoted by
$\mbox{\textup{drivers}}_g(z_0)$. $\Box$

\begin{thm}\label{II.004thm}

\begin{enumerate}

\item  The coefficient $e_g(-s)$ is of class iir in $z_0$ with
singularities restricted to $z_0=0$ and $z_0=\nu/(\nu-1)$.

\item  The solution of (\ref{II.007}) may be represented as
\begin{align} \label{egform}
e_g(-s) &=
 - \frac{1}{(2-2g)(1-2g)}
  \mbox{\textup{drivers}}_g(z_0(s))
 \\ \nonumber &  - \frac{1}{2-2g} \left( \frac{c_\nu
z_0(s)^\nu}{z_0(s)-1} \right)^{(2-2g)/(\nu-1)} \int_1^{z_0(s)}
\left( \frac{y-1}{c_\nu y^\nu} \right)^{(2-2g)/(\nu-1)} \left(
\mbox{\textup{drivers}}_g(y) \right)^{\bullet} dy  \\
\nonumber&  + \frac{1}{(1-2g)} \left( \frac{c_\nu
z_0(s)^\nu}{z_0(s)-1} \right)^{(1-2g)/(\nu-1)} \int_1^{z_0(s)}
\left( \frac{y-1}{c_\nu y^\nu} \right)^{(1-2g)/(\nu-1)}
\left(\mbox{\textup{drivers}}_g(y) \right)^{\bullet} dy
\\ \nonumber &
+ K_1 s^{(2g-2)/(\nu-1)} + K_2 s^{(2g-1)/(\nu-1)}
\end{align}
when $g \neq 1$, where $K_1$ and $K_2$ are constants of
integration either determined by the requirement that $e_g$ be a
locally analytic function of $s$ or by the evaluation of $e_g$ for
low values of $\nu$ through its combinatorial characterization;
and, when $g=1$,
\begin{align} \nonumber
e_1(-s) &= \frac{1}{(\nu-1)} \left[ \left( \frac{z_0(s)-1}{c_\nu
    z_0(s)^\nu} \right)^{1/(\nu-1)} \int_1^{z_0(s)} \left( \frac{c_\nu
    y^\nu}{y-1} \right)^{\nu/(\nu-1)} \frac{(\nu - (\nu-1) y)}{c_\nu
    y^{\nu+1} } \mbox{\textup{drivers}}_1(y) dy
\right. \\ \nonumber &\phantom{= \frac{1}{(\nu-1)} } \left.
-\int_1^{z_0(s)} \frac{ (\nu - (\nu-1) y)}{y (y-1)}
    \mbox{\textup{drivers}}_1(y) dy \right] \\
\label{e1form}
&= -\frac{1}{12} \log\left( \nu - (\nu-1) z_0(s) \right),
\end{align}
where we have chosen the principal branch of the logarithm.
By $\left(\mbox{\textup{drivers}}_g(y) \right)^{\bullet}$ we mean
the derivative of $\mbox{\textup{drivers}}_g(y)$ with respect to
$y$.
\end{enumerate}

\end{thm}

\section{Leading Order}
\label{e0_calc_section}

We will show that the leading order coefficient, $e_0(x,t)$, of the
asymptotic partition function,
$k^{-2}\log(Z_N^{(k)}(t)/Z_N^{(k)}(0))$ is found in terms of the
equilibrium measure $\mu = \mu_{V_t/x}$.
Define $\psi(\lambda)$ by
\begin{equation*}
d\mu = \psi(\lambda) d\lambda.
\end{equation*}
We denote the leading order behavior
\begin{equation*}
E_t = \lim_{k \to \infty} -\frac{1}{k^{2}} \log\left(Z_N^{(k)}(t)\right) =
I(x,t),
\end{equation*}
and
\begin{equation}\label{3.1}
\lim_{k \to \infty} -\frac{1}{k^{2}} \log\left(
\frac{Z_N^{(k)}(t)}{Z_N^{(k)}(0)}
  \right) =  E_t - E_0.
\end{equation}
More explicitly, by (\ref{eq:lead02})
\begin{eqnarray} \nonumber
E_t &=& \int \frac{V_t(\lambda)}{x} d\mu(\lambda) + \int \int
\log|\lambda - \eta|^{-1} d\mu (\lambda) d\mu(\eta)
\\ \label{inner}
&=& \left( \frac{ V_t(\lambda)}{x}, \psi(\lambda) \right) -
\left( \mathcal{L} \psi, \psi \right),
\end{eqnarray}
where \index{$\mathcal{L}$}
$$ \left( \mathcal{L} f \right)(\lambda)
= \int \log|\lambda - \eta| f(\eta) d\eta $$ is the logarithmic
potential of the measure $f(\eta)d\eta$, and where the inner
product $(\cdot, \cdot)$ is defined by
\begin{equation*}
(f, g) = \int f(\lambda) g(\lambda) d\lambda.
\end{equation*}
Using (\ref{3.1}) and (\ref{inner}) together with
\eqref{I.002}, we find that
\begin{equation}\label{e0}
e_0(t,x) = - \frac{1}{x}\left( \frac{1}{2} \lambda^2 + t
\lambda^{2\nu}, \psi(\lambda) \right) + \left(
\mathcal{L}\psi, \psi \right) + \frac{1}{x}\left( \frac{1}{2}
\lambda^2, \psi_0(\lambda) \right) -  \left( \mathcal{L} \psi_0,
\psi_0 \right).
\end{equation}

We recall here the parameter $l$ introduced in (\ref{Var}):
\begin{eqnarray*}
l &=& \int 2 \log| \lambda - s|^{-1} \psi(s) ds +
\frac{V(\lambda)}{x}
\\
&=& - 2 ( \mathcal{L} \psi, \psi) + \frac{V(\lambda)}{x} .
\end{eqnarray*}
Using this we have the following reduced formula for $e_0$:
\begin{equation} \label{3.4}
e_0 = -\frac{ (V, \psi)}{ 2 x} + \frac{(-l)}{2} + \frac{1}{4 x}
(\lambda^2, \psi_0)  - \frac{(-l_0)}{2}.
\end{equation}
This formula shows that there are two fundamental quantities that
need to be calculated in order to evaluate $e_0$. These are the
moment $(V, \psi)$ and the \emph{lagrange multiplier} $l$
associated to the constraint that the measure $\mu$ should have
total mass = 1. The other quantities appearing in (\ref{3.4}),
$(\lambda^2, \psi_0)$ and $l_0$ are evaluated by specializing the
fundamental quantities at $t=0$. Evaluating the fundamental
quantities will require an explicit asymptotic expansion, for
large $\lambda$, of the equilibrium measure
$\psi(\lambda)d\lambda$. We develop this in the next section.

\subsection{Explicit asymptotic expansion of the equilibrium measure}

The equilibrium measure $\mu$ is of the form
\begin{equation}
\psi(\lambda) = \frac{1}{2 \pi i} h(\lambda) \sqrt{\lambda^2 -
  \beta^2} \;\chi_{[-\beta, \beta]}(\lambda),
\end{equation}
where  $h$ is a polynomial determined by

\begin{equation} \label{5_3}
h(\lambda) = \frac{1}{2\pi i } \oint \frac{ V_t'(s)}{x \sqrt{s^2
- \beta^2} (s - \lambda) } ds,
\end{equation}
the integral in (\ref{5_3}) is taken along a simple closed
counterclockwise contour large enough to contain the interval
$[-\beta, \beta]$ and the point $\lambda$.

We will now evaluate the polynomial $h$ by expanding the
integrand of (\ref{5_3}) for large $\lambda$ and calculating the
loop integral on this expansion.

Define the sequence $\{ v_j\}_{j=0}^\infty$ by
\begin{equation} \index{$v_j$}
\frac{ \sqrt{\lambda^2 - \beta^2}}{\lambda} = 1 -
\sum_{i=0}^{\infty}
  v_i \frac{1}{\lambda^{2i+2}};
\end{equation}
whose Taylor coefficients can be computed to be
\begin{equation} \label{v_k}
v_i = \frac{1}{4^i} \binom{2i-1}{i-1} \frac{\beta^{2i+2}}{i+1},
\end{equation}
with $v_0$ defined to be $\beta^2/2$.

  Expand the polynomial $h$, given by (\ref{5_3}), in terms of its
coefficients $h_j$:
\begin{equation} \index{$h$!$h_j$}
h(\lambda) = \frac{1}{x} \left( 1 + \sum_{j=0}^{\nu - 1} h_j
\lambda^{2j} \right).
\end{equation}
Next note that
\begin{equation}
\frac{\lambda}{\sqrt{ \lambda^2 - \beta^2}} = \sum_{i=0}^\infty
\frac{
  2(i+1) v_i }{ \beta^2 } \frac{1}{\lambda^{2i} } .
\end{equation}
A direct computation of (\ref{5_3}) gives
\begin{equation} \label{h_j}
h_j = 4 \nu (\nu - j) t \frac{ v_{\nu -1 -j} }{\beta^2 }.
\end{equation}
The constraint that the total mass of $\mu$ should be 1,
\begin{equation} \label{4_12}
1 = \int_{-\beta}^\beta \psi(\lambda) d\lambda = \frac{1}{2} \oint
\psi(\lambda) d\lambda,
\end{equation}
implicitly determines $t$ as a function of $\beta^2$. Here, the
integral is over a contour containing the interval $[-\beta,
\beta]$.

We compute the loop integral in (\ref{4_12}) over a large contour
and find that the constraint can be expressed as a relation
between the coefficients $\{h_j\}$ and $\{v_j\}$:
\begin{equation} \label{4_13}
2 = \frac{ v_0}{x} + \sum_{j=0}^{\nu-1} \frac{h_j v_j}{x}.
\end{equation}
The relation (\ref{4_13}) simplifies  to the expression (using
identities (\ref{v_k}) and (\ref{h_j}) )
\begin{equation} \label{con1}
2 x = \frac{\beta^2}{2} + \frac{\nu}{4^{\nu-1}}
\binom{2\nu-1}{\nu-1}
 t \beta^{2\nu}.
\end{equation}
If we set $z=\beta^2/(4x)$ (\ref{con1}) becomes
\begin{equation} \label{con2}
1 = z + 2 \nu \binom{2\nu-1}{\nu-1} x^{\nu-1} t z^\nu.
\end{equation}
When $\nu=2$,
\begin{equation}
h(\lambda) = \frac{1}{x} ( 1 + 2 t \beta^2 + 4 t\lambda^2),
\end{equation}
with  the constraint
\begin{equation} \label{4_17}
1 = z + 12 x t z^2 .
\end{equation}

\subsection{Explicit calculation of $l$}

We will first derive an analytic expression for $l$ in terms of
$\beta^2$. To this end we study the logarithmic potential of
$\mu$:
\begin{equation} \index{$g(\lambda)$}
g(\lambda) = \int_{-\beta}^\beta \log( \lambda - s) \psi(s) ds.
\end{equation}
This function is analytic in $\mathbb{C}\setminus (-\infty,
\beta]$. For $\lambda \in (-\infty, \beta]$ we define two
functions $g_+$ and $g_-$ by
\begin{equation*}
g_\pm(\lambda) = \lim_{\epsilon \to 0} g(\lambda \pm i \epsilon).
\end{equation*}
Choosing the appropriate branch of the logarithm we find that
these functions are expressible as
\begin{equation} \label{4_20}
g_{\pm}(\lambda) = \int
 \log|\lambda - s|\psi(s) ds \pm i\pi
 \int_{\lambda}^\beta \psi(s) ds.
\end{equation}
A calculation using (\ref{4_20}) and (\ref{Var}) shows that
\begin{equation} \label{4_21}
g_+(\lambda) + g_-(\lambda) - \frac{V(\lambda)}{x} + l = 2 \int
\log|\lambda - s| \psi(s) ds - \frac{V(\lambda)}{x} + l = 0,
\end{equation}
and
\begin{equation} \label{4_22}
g_+(\lambda) - g_-(\lambda) = -2 i \pi \int_\beta^\lambda \psi(s)
ds.
\end{equation}
Equation (\ref{4_22}) implies that
\begin{equation*}
g_+(\lambda) = g_-(\lambda) - 2 i \pi \int_\beta^\lambda \psi(s)
ds;
\end{equation*}
therefore, if $\lambda$ is in $\mathbb{C}/(-\infty, \beta]$, then
equation (\ref{4_21}) becomes
\begin{equation*}
2 g(\lambda) - 2 i \pi \int_\beta^\lambda \psi(s) ds -
\frac{V(\lambda)}{x} + l = 0,
\end{equation*}
and
\begin{equation} \label{ell}
- l =  2 g(\lambda) - \frac{V(\lambda)}{x} + \int_\beta^\lambda
h(s) \sqrt{s^2 - \beta^2} ds.
\end{equation}
This is the basic expression we shall use to calculate $l$. Since
$l$ is a constant it can be evaluated for any choice of $\lambda$.
We will evaluate it by studying the limit of (\ref{ell}) as
$\lambda \to \infty$.  First observe that in this limit
\begin{equation}
g(\lambda) = \log(\lambda) +
\mathcal{O}\left(\frac{1}{\lambda}\right).
\end{equation}

Thus the principal issue is to develop an asymptotic expansion of
the indefinite integral
\begin{equation} \label{4_25}
 \int_\beta^\lambda h(s) \sqrt{ s^2 - \beta^2}ds.
\end{equation}
The details of the derivation are deferred to Appendix \ref{A};
the result is the following:
\begin{eqnarray} \label{intexp}
\int_\beta^\lambda h(s) \sqrt{s^2 - \beta^2} &=& \frac{1}{x} W(\lambda) \left(
\lambda^2 - \beta^2\right) \frac{\sqrt{\lambda^2 -
\beta^2}}{\lambda} + \frac{ 2 }{\beta^2} \lambda^2 \frac{
\sqrt{\lambda^2 -
    \beta^2}}{\lambda}
\\ \nonumber & &
- 2 \log\left( \frac{\lambda}{\beta} + \frac{\sqrt{\lambda^2 -
    \beta^2}}{\beta} \right),
\end{eqnarray}
where
\begin{equation} \index{$W(\lambda)$}
W(\lambda) = \sum_{p=1}^{\nu-1} w_p \lambda^{2p},
\end{equation}
with
\begin{equation} \index{$W(\lambda)$!$w_p$}
w_p = \frac{1}{2 v_p (p+1)} \sum_{j=p}^{\nu-1} h_j v_j.
\end{equation}

As a result of these calculations and taking the limit as $\lambda
\rightarrow \infty$, we deduce the formula
\begin{equation} \label{5_30}
-l = \beta^2 \sum_{j=1}^{\nu-1} w_j v_{j-1} - \sum_{j=1}^{\nu-1}
w_j v_j - 1 + \log\left( \frac{\beta^2}{4} \right),
\end{equation}
in which terms of non-constant order (which must cancel in any
case) have been dropped.
\bigskip

{\bf Remark}  The relations generated by setting the non-constant
terms equal to zero are equivalent to the moment conditions
\ref{thm:lead} (4) for the measure $\psi$ (see also
\cite{PierceThesis}). $\Box$

Equation (\ref{5_30}) simplifies to the expression:
\begin{equation} \label{4_37}
-l = \frac{ 4(2\nu-1)(2\nu-3)! t}{ x \nu (\nu-2)!^2 }
\frac{\beta^{2\nu}}{4^\nu} - 1 + \log\left( \frac{\beta^2}{4}
\right).
\end{equation}

Next, we use (\ref{con1}) to express $t$ as a function of
$\beta^2$ and $x$:
\begin{equation} \label{5_36}
t = -\frac{  (\nu-1)!^2 4^{\nu-1} }{2 (2\nu-1)! \beta^{2\nu}} (
\beta^2 - 4 x ).
\end{equation}
Substituting (\ref{5_36}) into (\ref{4_37}) we finally have:
\begin{equation} \label{4_39}
-l = -\frac{\nu-1}{\nu} \left( \frac{\beta^2}{4x} - 1 \right) -1 +
\log\left( \frac{\beta^2}{4} \right) .
\end{equation}
As an example, consider $\nu=2$ for which equation (\ref{4_39})
becomes
\begin{equation*}
-l = -\frac{1}{2} \left(\frac{\beta^2}{4x} - 1 \right) -1 +
\log\left(\frac{\beta^2}{4}\right).
\end{equation*}

\subsection{Explicit calculation of $(V, \psi)$}

Next we need to find an expression for $(V, \psi)$:
\begin{equation} \label{5_37}
\left( V, \psi \right) = \frac{1}{x} \int_{-\beta}^\beta
\left(\frac{1}{2} \lambda^2 + t \lambda^{2\nu} \right) h(\lambda)
\sqrt{\lambda^2 -
  \beta^2} d\lambda.
\end{equation}
Evaluating (\ref{5_37}) by regarding its double as a contour
integral and computing the value in terms of $v_j$ and $h_j$, we
find that
\begin{equation}
\left(V, \psi \right) = \frac{1}{4x} \sum_{j=0}^{\nu-1} h_j
v_{j+1} + \frac{v_1}{4x} + \frac{t}{2x} \sum_{j=0}^{\nu-1} h_j
v_{j+\nu} + \frac{
  t v_\nu}{2x}.
\end{equation}
This expression simplifies to
\begin{equation} \label{4_46}
\left(V, \psi \right) = \frac{ 4^{-1-\nu} t \beta^{2\nu+2} (2\nu)!
}{
  x (\nu+1) (\nu-1)!^2} + \frac{\beta^4}{ 32 x } + \frac{ 24^{-2\nu} t^2
  (2\nu-1)!^2 \beta^{4\nu}}{ x\nu (\nu-1)!^4 } + \frac{
  4^{-1-\nu}(\nu+1) t \beta^{2\nu+2} (2\nu)!}{x (\nu+1)!^2 }.
\end{equation}
As before, we simplify equation (\ref{4_46}) by substituting the
expression (\ref{5_36}) for $t$ in terms of $\beta^2$ :
\begin{equation}
\left( V, \psi \right) = -\frac{ -8 x \beta^2 \nu^2 + \beta^4
\nu^2 -
  2\beta^4 \nu - x^2 16 \nu - x^2 16 + 8\beta^2 \nu x + \beta^4 }{  32 x (\nu+1)\nu}.
\end{equation}
When $\nu=2$, we find that
\begin{equation}
\left( V, \psi \right) = \frac{1}{12 } \beta^2 - \frac{1}{ 192
x} \beta^4 + \frac{x}{4}.
\end{equation}

\subsection{The explicit formula for $e_0(t)$}

We can now put all of the pieces together.  Evaluating the above
 expressions when
 $t = 0$ (or equivalently when $ \beta = 2\sqrt{x}$),
we find $E_0 = -3/4 + 1/2 \log(x)$.  Collecting all the components
gives
\begin{equation} \label{fact}
e_0 = \frac{1}{16 x^2} \mu \left( \beta^2 - 4x \right)\left(
\beta^2 - 4x r \right) + \frac{1}{2} \log\left( \frac{\beta^2}{4x}
\right) = \eta \left( z-1\right)\left(z-r\right)+\frac{1}{2}
\log(z),
\end{equation}
where
\begin{eqnarray} \label{mu}
\eta &=& \frac{ (\nu-1)^2 }{ 4 \nu (\nu+1)}, \\
\label{r} r &=& \frac{ 3 (\nu+1) }{\nu-1},
\end{eqnarray}
and
\begin{equation}
\label{z} z = \frac{\beta^2}{4x}.
\end{equation}
We have proven that $e_0$ has an explicit representation depending
only on $\nu$ (see the first part of Theorem \ref{II.001}).

When $\nu=2$ the result is
\begin{equation}
e_0 = \frac{1}{24} (z-1)(z-9) + \frac{1}{2} \log(z).
\end{equation}
\bigskip

{\bf Remark}  It is interesting to compare our expression here to
a formula derived in \cite{BI}.  The difference is that while our
formula is more explicit, it is restricted to a 1-parameter family
of times corresponding to a fixed valence of the vertices. $\Box$

\subsection{Enumeration of Planar Graphs}

It follows from Theorem \ref{thm:01} that the coefficients,
$\kappa_0^{(\nu)}(j)$ , of the leading order term $e_0(t)$ count
the number of planar $2\nu$-regular maps with $j$ vertices. We
have shown that $e_0(t)$ can be explicitly expressed in terms of
the auxiliary function $z(t)$. The latter solves the algebraic
relation (\ref{con2}).  So to find the Taylor Coefficients of
$e_0(t)$, we must first find the Taylor Coefficients of $z(t)$.

We define
\begin{equation}\label{alpha}
 \alpha = - c_\nu x^{\nu-1} t,
\end{equation}
 where
$$
c_\nu = 2\nu \left(\begin{array}{c}
  2\nu-1\\
  \nu-1\\
\end{array}\right).
$$
Then the polynomial relation defining $z(t)$ is
\begin{equation} \label{6_0}
1 = z(t) - \alpha z(t)^\nu
\end{equation}
in a neighborhood of $t = 0$. (We view $x$ here as a scaling
parameter. When $x$ is set to 1, we will recover the counting
function.)

The $j^{th}$ coefficient of the Taylor expansion of $z$ as a
function of $\alpha$ near $0$, $z = \sum_{j\geq 0} \zeta_j
\alpha^j$, is of course given by

$$
\zeta_j = \frac{1}{2\pi i} \oint \frac{z(\alpha)}{\alpha^{j+1}}
d\alpha.
$$
Making the substitution $u=z(\alpha)$, and using the evaluation
\begin{equation}\label{dreln}
\frac{dz}{d\alpha} =\frac{z^{\nu}}{1-\nu\alpha z^{\nu -1}}
\end{equation}
derived by differentiating the relation (\ref{6_0}), and also
using (\ref{6_0}) to eliminate $\alpha$, this integral becomes
\begin{equation}\label{loop}
\frac{1}{2\pi i}\oint_{u\sim 1} \frac{\left(\nu -
(\nu-1)u\right)u^{\nu j}}{(u-1)^{j+1}} du.
\end{equation}
Applying the binomial expansion, it is then straightforward to
evaluate this loop integral and find that
\begin{equation}\label{zcoeff}
\zeta_j = (1-\nu)\left(\begin{array}{c}
  \nu j +1\\
  j\\
\end{array} \right)  + \nu \left(\begin{array}{c}
  \nu j \\
  j\\
\end{array}\right)  = \frac{1}{j}\left(\begin{array}{c}
  \nu j \\
  j-1 \\
\end{array}\right).
\end{equation}

We note that these coefficients are precisely the \emph{higher
Catalan numbers}, which play a role in a wide variety of
combinatorial problems. For a discussion of these applications and
their relation to the work discussed here see \cite{Pierce}.

A similar approach will yield coefficients for the other terms in
the expression (\ref{fact}) for $e_0$. For instance the Taylor
coefficients of $\log(z(\alpha))= \sum_{j=1}^\infty L_j^{(\nu)}
\alpha^j$ are given by
\begin{eqnarray*}
  L_j^{(\nu)} &=& \frac{1}{2\pi i} \oint_{u\sim 1} \frac{\log(u)(1-\nu\alpha
u^{\nu-1})u^{j\nu}}{(u-1)^{j+1}} du.
\end{eqnarray*}
Expanding the integrand in the vicinity of $u=1$, we can evaluate
these coefficients as
\begin{equation}\label{logcoeff}
L_j^{(\nu)} = \sum_{k+\ell=j-1}\left\{
(1-\nu)\left(\begin{array}{c}
  \nu j \\
  j\\
\end{array} \right)  + \nu \left(\begin{array}{c}
  \nu j -1\\
  j\\
\end{array}\right) \right\} \frac{(-1)^k}{k+1} = \frac{1}{j}\left(\begin{array}{c}
  \nu j -1\\
  j-1 \\
\end{array}\right)
\end{equation}
By the same method one also derives the coefficients in the
expansion of the quadratic term, $(z(\alpha)-1)^2 =
\sum_{j=2}^\infty U_{2,j}^{(\nu)} \alpha^j$:
\begin{equation} \index{$U_{2,j}^{(\nu)}$}
U_{2,j}^{(\nu)} = \frac{2}{j} \binom{\nu j }{j-2}.
\end{equation}
Finally one may assemble all these contributions in (\ref{fact})
and substitute for $\alpha$ as in (\ref{alpha}) to conclude that
the Taylor coefficients of $e_0$ with respect to $t_{2\nu}$
satisfy
\begin{eqnarray*}
\frac{\kappa_0^{(\nu)}(j)}{c_\nu^j} &=& j! \left[ - (r-1)\eta
\zeta_j + \eta U_{2,j}^{(\nu)} + \frac{1}{2} L_j^{(\nu)}
\right]\\
&=& \left( -  (r-1)\eta \frac{ (\nu j)!}{\left(
  (\nu-1)j + 1\right)!} +  \eta \frac{ 2(j-1)(\nu j)! }{ \left( (\nu-1)j
  + 2 \right)! } + \frac{1}{2} \frac{ (\nu j - 1)! }{ \left( (\nu-1)j
  + 2 \right)! } \right).
\end{eqnarray*}
After simplifying this equation we get
\begin{equation}
\kappa_0^{(\nu)}(j) = c_\nu^j \frac{ (\nu j - 1)! }{ \left(
(\nu-1)j + 2
  \right)! }.
\end{equation}
This establishes the second half of Theorem \ref{II.001}.

  \section{Continuum Toda Equations}

  We will show that the Toda Lattice equations at level $2\nu$,
with the initial conditions $b_k = \sqrt{k}$, possess a continuum
limit under an appropriate scaling. These equations will lead to a
description of the evolution of the asymptotic partition function
coefficients, $e_g(t)$.  More precisely, in Section 4.1 we
establish Theorem 2.2 and in subsection 4.2 we establish Theorem 2.7.
To this end, we first review some basic facts about the Toda lattice
equations.

The Toda Lattice Hierarchy is given by the differential equations
\begin{equation} \label{Toda}
\frac{dL}{d\xi_{i}} = \left[ B_{i}(L), L\right],
\end{equation}
where $L$ is a symmetric tridiagonal matrix of the form
\begin{equation}
L =
\begin{pmatrix}
a_1   & b_1 & 0   & 0   & 0   & ... \\
b_1 & a_2   & b_2 & 0   & 0   & ... \\
0   & b_2 & a_3  & b_3 & 0   & ... \\
0   & 0   & b_3 & a_4  & b_4 & ... \\
\vdots & \vdots & \vdots & \vdots & \vdots&
\end{pmatrix},
\end{equation}
and $B_{i}(L) = \left( L^{i} \right)_+ - \left( L^{i} \right)_- $
where the plus subscript denotes upper triangular projection and
the minus subscript denotes lower triangular projection.
\bigskip

We briefly recall the notion of \emph{tau functions} in terms of
which the solution of the Toda Lattice equations at different time
parameters can be expressed. Let $L_0$ be a semi-infinite
tri-diagonal matrix of the type that arises as a 3-term recurrence
relation for orthogonal polynomials with exponential weight. The
dependence of the $j^{th}$ tau function on the $\xi_i$ is given by
\begin{equation}  \label{tau1} \index{$\tau_j(t)$}
\tau_j(\xi_1,\ldots,\xi_\upsilon;L_0) =  \left( \det\left[ M_j^2
\right] \right)^{1/2},
\end{equation}
where $M = \exp( \sum_{i=1}^\upsilon \xi_i L_0^i)$ and $M_j$ is the
$j \times j$ upper left block of $M$ in the basis of orthogonal
polynomials with respect to $d\mu$.  From the matrix factorization
method of solving the Toda equations \cite{Flaschka} one finds
that
\begin{equation} \label{tau_solution}
b_j(\vec{\xi}\ ) = \frac{ \tau_{j+1}(\vec{\xi}\ )
\tau_{j-1}(\vec{\xi}\ )}{\tau_j(\vec{\xi}\ )^2} b_j(0).
\end{equation}
Furthermore, from (\ref{tau_solution}) and the differential
equations for the $\xi_1$-Toda flow, one can deduce that
\begin{equation} \label{7_13}
b_j(\vec{\xi}\ ) = \sqrt{ \frac{1}{2} \frac{d^2}{d\xi_1^2} \log\left(
\tau_j(\vec{\xi}\ )
  \right) }.
\end{equation}
The representations (\ref{tau_solution}) and (\ref{7_13}) of $b_j$
in terms of tau functions are referred to as \emph{Hirota
relations}, familiar from the theory of soliton equations.

Finally, through the use of Hankel determinants, the tau function may
be related to the partition function, $Z_k = Z_k^{(k)}$.  Recall that
\cite{Szego} the Hankel determinant
associated to the weight $d \omega(\lambda) =
\exp( \sum_{i=1}^\upsilon \xi_i \lambda^i) d \lambda $ is defined as follows
\begin{eqnarray}
H_{k}(\vec{\xi}\ ) =  \ \left|
\begin{array}{cccc}
c_{0} & c_{1} &\cdots & c_{k-1} \\
c_{1} & c_{2} & \cdots & c_{k} \\
\vdots & \vdots &\ddots & \vdots \\
\vdots & \vdots & \ddots &\vdots \\
c_{k-1}& c_{k} & \cdots & c_{2 k - 2} \\
\end{array} \right|,
\end{eqnarray}
where $c_{j} = \int_{\mathbb{R}} \lambda^{j} d \omega(\lambda)$ are the
moments of $d \omega$.  There follows the relation
\begin{equation} \nonumber
b_j^{2}(\vec{\xi}\ ) = \frac{ H_{j+1}(\vec{\xi}\ )
H_{j-1}(\vec{\xi}\ )}{H_j(\vec{\xi}\ )^2}.
\end{equation}
The partition function may be obtained from the Hankel determinant by
a rescaling of the time parameters (see, for example, \cite[Equation
(1.29)]{EM}).   Consequently one can relate the partition
function to the tau function, in the case of interest to us, as
\begin{equation} \label{tau_v1}
\tau_k(\xi_1, \xi_{2\nu}; L_0)^2 = \frac{Z_k\left( -\xi_1 ,-2
k^{\nu-1} \xi_{2\nu}  \right)}{Z_k(0)},
\end{equation}
where $L_0$ is the semi-infinite tridiagonal matrix with $b_k =
\sqrt{k}$, $a_k \equiv 0$, and where $\xi=\xi_{2\nu}$. For this
initial condition, the evolution of $L$ under the even $2\nu$-flow
preserves the vanishing of $a_k$. It will be natural to study this
Toda evolution in terms of the self-similar variable
\begin{equation}
s = 2 k^{\nu-1} \xi.
\end{equation}

We will start from the {\it unscaled} partition function
(\ref{I.001}):
\begin{eqnarray} \label{unscaled}
& & Z_1^{(k)}(-\xi_1, -\xi) =\\
\nonumber & & \int \cdots \int \exp\left\{ -\sum_{j=1}^{k}
\left(\frac{1}{2}\lambda_j^2 -\xi \lambda_j^{2\nu}
- \xi_1 \lambda_j\right) \right\} {\mathcal{V}} (\lambda) \,\, d^{k} \lambda, \\
\nonumber & & \mbox{where} \,\,\, \mathcal{V}(\lambda)=
\prod_{j\neq \ell} {| \lambda_{j} - \lambda_{\ell} | }.
\end{eqnarray}

Using the Hirota relations, (\ref{tau_solution}) and (\ref{7_13}),
we have the following Proposition.

\begin{prop} \label{prop_8}

The Toda recurrence coefficients may be represented as follows.
\begin{equation} \label{Z_k_is_tau}
b_k^2 = \frac{1}{2k} \frac{d^2}{d{t_1}^2} \log\left[ Z_k(
{t_1},-s) \right] \bigg|_{t_1 = 0}, \quad \mbox{where} \quad s\leq
0.
\end{equation}
In addition, the following asymptotic expansion holds true.
\begin{equation} \index{$z_g(s)$}
b_k^2 = k \sum_{g \geq 0} z_g(s) \frac{1}{k^{2g}},
\end{equation}
where $z_g(s)$ is an analytic function of $s$ in a neighborhood of
$s=0$, and
\begin{equation}
z_0(0) = 1, \, z_g(0) = 0, \, g > 0.
\end{equation}

\end{prop}

To prove this proposition we scale the eigenvalues in the unscaled
partition function as $\lambda_j = N^{1/2}\tilde{\lambda}_j$ under
which

\begin{eqnarray}
 Z_1^{(k)} &=& N^{k^2/2}\int \cdots \int
  \exp\left(-N \left\{\sum_{j=1}^{k} \left(\frac{1}{2}\tilde{\lambda}_j^2
-\xi N^{\nu-1}\tilde{\lambda}_j^{2\nu} - \frac{\xi_1}{\sqrt{N}}
\tilde{\lambda}_j\right) \right\}\right)
  {\mathcal{V}} (\tilde{\lambda}) \,\, d^{k} \tilde{\lambda}
\nonumber
\\
 &=&
N^{k^2/2} Z_N^{(k)}\left( -\frac{\xi_1}{\sqrt{N}},-\xi
N^{\nu-1}\right).
\end{eqnarray}

We observe that the Toda variables associated to our random matrix
ensemble with the corresponding N scaling have the following
representation:

\begin{eqnarray}
b_k^2 &=& \frac{1}{2}\frac{d^2}{d\xi_1^2}\log Z_1^{(k)}\\
\nonumber  &=& \frac{1}{2}\frac{d^2}{d\xi_1^2} \log \left[
Z_N^{(k)}\left(-\frac{\xi_1}{\sqrt{N}},-\xi N^{\nu-1}
            \right)\right]\\
\nonumber  &=& \frac{1}{2N}\frac{d^2}{d{t_1}^2} \log \left[
Z_N^{(k)}({t_1}, {t})\right]\bigg|_{{t}=-2\xi N^{\nu-1},
{t_1}=-\frac{\xi_1}{\sqrt{N}}}.
\end{eqnarray}

We next define a spatial scale through the ratio between $k$ and
$N$: set $k=xN$. If we furthermore introduce the scaling
$\tilde{\lambda}_j = \sqrt{x}\hat{\lambda}_j$, then we can rewrite

\begin{eqnarray}  \nonumber
 Z_N^{(k)}({t_1}, {t}) &=&
x^{k^2/2}\int \cdots \int
  \exp\left(-k \left\{\sum_{j=1}^{k} \left(\frac{1}{2}\hat{\lambda}_j^2
+ {t} x^{\nu-1}\hat{\lambda}_j^{2\nu} + \frac{{t_1}}{\sqrt{x}}
\hat{\lambda}_j\right) \right\}\right)
  {\mathcal{V}} (\hat{\lambda}) \,\, d^{k} \hat{\lambda}\\
&=& x^{k^2/2} Z_k^{(k)}\left(\frac{{t_1}}{\sqrt{x}},{t} x^{\nu-1}
\right).
\end{eqnarray}

Putting all the above together, the natural variable to consider
for this continuum limit of the Toda Lattice Equations is
\begin{eqnarray} \label{9_14}
\frac{1}{k}b_k^2 = \frac{1}{2k^2}\frac{d^2}{d{t_1}^2} \log \left[
Z_k^{(k)}\left({t_1},-s
\right)\right]\bigg|_{{t_1}=-\frac{1}{\sqrt{k}}\xi_1,s = 2
k^{\nu-1}\xi}.
\end{eqnarray}
One may conclude directly from the main theorem of \cite{EM} that
$\frac{1}{k^2}\log Z_k^{(k)}$ has a full asymptotic expansion in
powers of $k^{-2}$, uniformly valid in an appropriate $(t_1,-s)$
domain which includes arbitrary negative $s$ and simultaneously an
open neighborhood of $t_1=0$. On this domain the asymptotic
expansion of derivatives may
be calculated via term by term differentiation of the original
asymptotic expansion. Moreover, the coefficients in this expansion
are complex analytic in a neighborhood of $(0,0)$. It follows that

\begin{eqnarray}
\frac{1}{2k^2}\frac{d^2}{d{t_1}^2} \log \left[
Z_k^{(k)}\left({t_1},-s \right)\right]\bigg|_{{t_1}= 0}
\end{eqnarray}
has a full asymptotic expansion in powers of $k^{-2}$ uniformly
valid for negative $s$ and whose coefficients are analytic in a
complex neighborhood of $s=0$.

\subsection{The Continuum Toda Hierarchy}

In this subsection we prove Theorem 2.2.

We begin with the Toda lattice systems (\ref{Toda}) introduced
earlier. It will prove convenient to transform this to the
so-called Kostant-Toda lattice form \cite{EFS} where L is
asymmetric tri-diagonal but in which the $b_k^2$ are the
fundamental variables. This is achieved by conjugating $L$ in
(\ref{Toda}) by the diagonal matrix
\begin{equation}\label{D}
D = \begin{pmatrix}
1  & 0 & 0   & 0   & 0   & ... \\
0 &  b_1^{-1}  & 0 & 0   & 0   & ... \\
0   & 0 & (b_1 b_2)^{-1}   & 0 & 0   & ... \\
0   & 0   & 0 & (b_1 b_2 b_3)^{-1}  &  & ... \\
\vdots & \vdots & \vdots & \vdots & \ddots&
\end{pmatrix},
\end{equation}

The fundamental matrix variable for the Kostant-Toda system is
\begin{equation}\label{K-T}
\mathcal{L} = D^{-1}LD = \begin{pmatrix}
0   & 1 & 0   & 0   & 0   & ... \\
b_1^2 & 0   & 1 & 0   & 0   & ... \\
0   & b_2^2 & 0   & 1 & 0   & ... \\
0   & 0   & b_3^2 & 0   & 1 & \ddots \\
\vdots & \vdots & \ddots & \ddots & \ddots&
\end{pmatrix}.
\end{equation}
Because $D$ is diagonal it is straightforward to see that the RHS
of (\ref{Toda}) transforms as
\begin{equation}\label{Lax_R}
D^{-1}\left[ B_{2\nu}(L), L\right]D = \left[ D^{-1}B_{2\nu}(L)D,
D^{-1}LD\right] = \left[ B_{2\nu}(\mathcal{L}),
\mathcal{L}\right].
\end{equation}
One also has
$$
\mathcal{L}_{\xi} = (D^{-1}LD)_{\xi} = D^{-1}L_{\xi}D + [\mathcal{L},
D^{-1}D_{\xi}].
$$
Combining this observation with (\ref{Lax_R}) in (\ref{Toda}) we
see that the Lax equation transforms to
\begin{equation}\label{Lax}
\mathcal{L}_{\xi} + [D^{-1}D_{\xi}, \mathcal{L}] = \left[
B_{2\nu}(\mathcal{L}), \mathcal{L}\right].
\end{equation}
By direct calculation one can check that
$$
[D^{-1}D_{\xi}, \mathcal{L}] = -\frac{1}{2} \mathcal{L}_{\xi} + [D^{-1}D_{\xi},
\epsilon],
$$
where
\begin{equation}\label{epsilon}
\epsilon = \begin{pmatrix}
0   & 1 & 0   & 0   & 0   & ... \\
0 & 0   & 1 & 0   & 0   & ... \\
0   & 0 & 0   & 1 & 0   & ... \\
0   & 0   & 0 & 0   & 1 & \ddots \\
\vdots & \vdots & \ddots & \ddots & \ddots&
\end{pmatrix}.
\end{equation}
Using this observation the Lax equation (\ref{Lax}) becomes
\begin{equation}\label{Lax2}
\frac{1}{2} \mathcal{L}_{\xi} = \left[ B_{2\nu}(\mathcal{L}),
\mathcal{L}\right] - [D^{-1}D_{\xi}, \epsilon].
\end{equation}
Since the LHS is lower triangular, we may reduce this
equation to
\begin{equation}\label{Lax3}
\frac{1}{2} \mathcal{L}_{\xi} = \left[ B_{2\nu}(\mathcal{L}),
\mathcal{L}\right]_-.
\end{equation}
Since $\mathcal{L}$ is tri-diagonal, the RHS may be further
simplified as
$$
\left[ B_{2\nu}(\mathcal{L}), \mathcal{L}\right]_- = \left[
\mathcal{L}^{2\nu}_+ - \mathcal{L}^{2\nu}_-, \mathcal{L}\right]_-
= - \left[\mathcal{L}^{2\nu}_-, \mathcal{L}\right]_-.
$$
Finally, since $\mathcal{L} = \mathcal{L}_- + \epsilon$ with
$\mathcal{L}_-$ supported only on the first lower diagonal, it
follows that $\mathcal{L}_{\xi}$ is only supported on the first lower
diagonal and therefore we may use the following final form of our
Lax equation:
\begin{equation}\label{Laxf}
\frac{1}{2} \mathcal{L}_{\xi} = \left[\epsilon, \mathcal{L}^{2\nu}_-
\right]_{(-1)},
\end{equation}
where the subscript $(-1)$ denotes projection onto the first lower
subdiagonal. At the level of the matrix entries $b_k^2$ this
system of equations becomes
\begin{equation}\label{beqns2}
\frac{1}{2} \frac{db_k^2}{d\xi} = (\mathcal{L}^{2\nu})_{k+2,k} -
(\mathcal{L}^{2\nu})_{k+1,k-1}
\end{equation}
where,
\begin{equation}\label{beqns3}
(\mathcal{L}^{2\nu})_{k+1,k-1} = \sum_{i_1,i_2, \ldots, i_{2\nu+1};
 |i_{j+1}-i_j|=1; i_1=k+1, i_{2\nu+1}=k-1} \mathcal{L}_{k+1,i_2}\mathcal{L}_{i_2,i_3}\ldots \mathcal{L}_{i_{2\nu},k-1}.
\end{equation}
as stated in (\ref{beqns1}) and (\ref{beqns1.1}) .

Note that in (\ref{beqns3}) the sum may be viewed as being taken
over walks, $w$, from k+1 to k-1 of length $2\nu$. We will
sometimes represent a walk $w$ as a $2 \nu$-vector of $\pm 1's$.
It is clear that such a walk has $\nu + 1$ "downturns" (vector
entries of -1) and $\nu-1$ "upturns" (vector entries of +1).
Therefore, specifying the locations of the downturns determines
$w$. From this we see that the set of all such walks, $\{w\}$ is
in one-to-one correspondence with all choices of $\nu +1$ numbers
(the downturn locations), $\{j_1 < j_2 < \dots <j_{\nu +1}\}$ from
$\{1,\dots, 2\nu\}$:
$$
\sharp \{w\} = \left( \begin{array}{c}
  2\nu \\
  \nu +1 \\
\end{array}\right).
$$
Thus (\ref{beqns3}) becomes
\begin{equation}\label{beqns4}
    \sum_{\{w\}} b^2_{k+\ell_1(w)}b^2_{k+\ell_2(w)}\cdots
    b^2_{k+\ell_{\nu+1}(w)},
\end{equation}
where
\begin{equation}\label{beqns5}
    \ell_m(w) = j_m - 2m +1.
\end{equation}
Some explanation is perhaps necessary here.  The sum appearing in
(\ref{beqns4}) is taken over all choices of $\nu+1$ numbers, and
each such choice is denoted by a member $w$ of the set $\{w\}$. As
described above, each $w$ may be interpreted as a walk on the
integer lattice, from k+1 to k-1 of length $2 \nu$. Since the
structure of the walks and all associated counting is independent
of the value of k, we will from now on, for simplicity, take k=0.
The quantity $\ell_{m}(w)$ is defined to be the location on the
integer lattice after the $m$th downturn. It is now
straightforward to compute that, given a walk with downturns at
$j_{1}, \ldots, j_{\nu+1}$, after $m$ downturns, there have been
$j_{m}-m$ upward steps, and $m$ downward steps. Since the starting
position is +1, (\ref{beqns5}) follows.

Inserting these reductions into the Toda equations (\ref{beqns2})
we have
\begin{equation}\label{beqns6}
    \frac{1}{2}\left(b^2_k\right)_{\xi} = \sum_{\{w\}}\left[\prod_{m=1}^{\nu+1}b^2_{k+\ell_m(w)+1} -
    \prod_{m=1}^{\nu+1}b^2_{k+\ell_m(w)}\right].
\end{equation}

By Proposition \ref{prop_8},
\begin{eqnarray}
\label{beqns7}
  b^2_k(\xi)=b^2_k(\frac{s_k}{2k^{\nu-1}})=\widehat{b}^2_k(s_k) &=&
  k\left(\frac{1}{2k^2}\left.\frac{d^2}{dt_1^2}\log(Z_k({t_1},-s_k))\right|_{t_1=0}\right),
\end{eqnarray}
and the differential equation (\ref{beqns6}) becomes
\begin{eqnarray}
\label{beqns8}
  k^{\nu-1} \frac{d}{ds_{k}} \widehat{b}^{2}_{k}(s_{k}) =
\sum_{\{w\}}\left[\prod_{m=1}^{\nu+1}\widehat{b}^2_{k+\ell_m(w)+1}
(s_{k+\ell_{m}+1}) -
    \prod_{m=1}^{\nu+1}\widehat{b}^2_{k+\ell_m(w)}(s_{k+\ell_{m}})\right].
\end{eqnarray}

>From (\ref{beqns7}), and the asymptotic expansion satisfied by the
partition function appearing therein, one sees that for any fixed
integer $\ell$, the quantity $(k+\ell)^{-1}
\widehat{b}^{2}_{k+\ell}(s_{k+\ell})$ possesses an asymptotic
expansion in even inverse powers of $(k+\ell)$, which we express
in the following form:
\begin{eqnarray}
\label{HCAAA01} \widehat{b}^{2}_{k+\ell}(s_{k+\ell}) = k
\sum_{g\ge 0} z_{g}\left( s_{k} \left( 1 + \frac{\ell}{k}
\right)^{\nu-1} \right)  \left(  1 + \frac{\ell}{k} \right)^{1-
2g}k^{-2g}.
\end{eqnarray}
\bigskip

{\bf Fundamental Asymptotic Principle:}  Recall that
\begin{description}
    \item[i] the
asymptotic expansion satisfied by the partition function is
differentiable term by term
    \item[ii] the coefficients
$z_{g}(\bullet)$ are analytic functions of $\bullet$.
\end{description}
Therefore, the terms appearing in the summation in (\ref{HCAAA01})
may be Taylor expanded (note that $\ell/k$ is asymptotically
small) and the resulting multiple series may be re-summed.  One may
convince oneself of this as follows.  One truncates the expansion
(\ref{HCAAA01}) at order $k^{1-2h}$, and observes that the error
is $O(k^{-1-2h})$. Then one computes Taylor expansions to order
$k^{-2h}$ of all quantities appearing in the (now finite) sum, and
observes that the error is $O(k^{-1-2h})$.
\medskip

Following this principle, we now introduce the scheme $f(s,w)$:

\begin{eqnarray}\label{HCAAA02}
  f(s,w) &=&  \sum_{g\ge 0} z_{g}\left(
s_{k} w^{\nu-1}\right)  w^{1- 2g}k^{-2g}.
\end{eqnarray}
Using this scheme, the differential equation (\ref{beqns8})
becomes:
\begin{eqnarray}
\nonumber \left. \frac{d}{ds}f(s,w)\right|_{w=1} &=& \sum_{\{w\}} kf^{\nu+1}\left\{\prod_{m=1}^{\nu+1}\left[ 1 + \frac{f_w}{f}\left(\frac{\ell_m +1}{k}\right) + \frac{f_{w^{(2)}}}{2f}\left(\frac{\ell_m +1}{k}\right)^2 + \cdots + \frac{f_{w^{(h)}}}{h! f}\left(\frac{\ell_m +1}{k}\right)^h + \cdots \right]\right. \\
 && \hspace{0.59in} - \left. \prod_{m=1}^{\nu+1}\left[ 1 + \frac{f_w}{f}\frac{\ell_m}{k} + \frac{f_{w^{(2)}}}{2f}\left(\frac{\ell_m}{k}\right)^2 + \cdots + \frac{f_{w^{(h)}}}{h! f}\left(\frac{\ell_m}{k}\right)^h + \cdots
   \right]\right\} \label{HCAAA03}
   \\
 \nonumber &=&\sum_{\{w\}} \left\{(\nu+1)f^\nu f_w +
 \mathcal{O}(\frac{1}{k^2})\right\}\\
 \nonumber &=&(\nu+1)\left(\begin{array}{c}
   2\nu \\
   \nu+1\\
 \end{array}\right)f^\nu f_w +
 \mathcal{O}(\frac{1}{k^2}),
\end{eqnarray}
where we note that $(\nu+1)\left(\begin{array}{c}
   2\nu \\
   \nu+1\\
 \end{array}\right) = 2\nu\left(\begin{array}{c}
   2\nu -1\\
   \nu-1\\
 \end{array}\right) = c_\nu$ as claimed in the statement of
 Theorem \ref{II.002}.

 From the expansion (\ref{HCAAA02}) one may now confirm
 that the higher order terms in the expansion of the RHS have the
 form described in Theorem \ref{II.002}. The justification for the form of the
 coefficient $d_V^{(\nu,g)}$ stated there requires some further explanation
 which we now provide.

 Given the explicit determination (\ref{beqns5}) of the $\ell_m$
 in terms of the walks, it is natural to try to calculate
 $d_V^{(\nu,g)}$ by expressing this combinatorial quantity in
 terms of expectations with respect to the induced probability
 distribution on the family of random walks $\{w\}$ conditioned to
 begin at +1 and end at -1. To this end we observe that given an
 ordered sequence of numbers $m_1 < m_2 < \dots < m_{r} $
 selected from the set $\{1, \dots, \nu+1\}$, and another ordered sequence $i_1 < \dots < i_r$
 selected from $\{1, \dots, 2\nu\}$,
 \begin{eqnarray}\label{jointdist}
   &&\sharp\left\{w|\,\, m_j^{th} \,\,\mbox{downturn of}\,\, w \,\,\mbox{occurs at position}\,\, i_j, \,\, j=1,\dots, r\right\} \\
   \nonumber &=&
   \left(\begin{array}{c}
     i_1-1 \\
     m_1-1 \\
   \end{array}\right)\left(\begin{array}{c}
     i_2-i_1-1 \\
     m_2-m_1-1 \\
   \end{array}\right)\cdots\left(\begin{array}{c}
     i_r-i_{r-1}-1 \\
     m_r-m_{r-1}-1 \\
   \end{array}\right)\left(\begin{array}{c}
     2\nu-i_r \\
     \nu+1-m_r \\
   \end{array}\right).
 \end{eqnarray}
These joint probability distributions are fundamental for the
count we will now describe.

One sees directly from expanding the products in (\ref{HCAAA03}),
that
\begin{eqnarray}
 \nonumber d_V^{(\nu,g)} &=& \sum_{\{w\}}\frac{1}{\prod_{j=1}^{2g+1}r_j(V)!}\sum_{\begin{array}{c}
    m_1,\dots, m_{\rho(V)} \\
    \mbox{distinct}\\
    \in \{1,\dots,\nu+1\}\\
  \end{array}} \left(\prod_{i=1}^{\rho(V)}\left(\ell_{m_i}+1\right)^{|V_i|}-\prod_{i=1}^{\rho(V)}\left(\ell_{m_i}\right)^{|V_i|}\right)\\
 \nonumber  &=& \frac{1}{\prod_{j=1}^{2g+1}r_j(V)!}\sum_{\begin{array}{c}
    m_1,\dots, m_{\rho(V)} \\
    \mbox{distinct}\\
    \in \{1,\dots,\nu+1\}\\
  \end{array}} \sum_{\{w\}} \left(\prod_{i=1}^{\rho(V)}\left(\ell_{m_i}+1\right)^{|V_i|}-\prod_{i=1}^{\rho(V)}\left(\ell_{m_i}\right)^{|V_i|}\right) \\
\nonumber  &=& \frac{1}{\prod_{j=1}^{2g+1}r_j(V)!}\sum_{\sigma \in
\mathcal{S}_{\rho(V)}}\sum_{\begin{array}{c}
    m_1< \dots < m_{\rho(V)} \\
\in \{1,\dots,\nu+1\}\\
  \end{array}} \sum_{\{w\}}
  \left(\prod_{i=1}^{\rho(V)}\left(\ell_{m_{\sigma(i)}}+1\right)^{|V_i|}-\prod_{i=1}^{\rho(V)}\left(\ell_{m_{\sigma(i)}}\right)^{|V_i|}\right)\\
\nonumber &=& \frac{1}{\prod_{j=1}^{2g+1}r_j(V)!}\sum_{\sigma \in
\mathcal{S}_{\rho(V)}} \sum_{1 \leq i_1 \dots < i_{\rho(V)} \leq
2\nu} \sum_{\begin{array}{c}
    m_1 < \dots < m_{\rho(V)} \\
\in \{1,\dots,\nu+1\}\\
  \end{array}} \\
\nonumber && \hspace{-.8in} \left(\prod_{j=1}^{\rho(V)}\left(i_{j} -
2m_{j}+2\right)^{|V_{\sigma(j)}|}\left(\begin{array}{c}
     i_1-1 \\
     m_1-1 \\
   \end{array}\right)\left(\begin{array}{c}
     i_2-i_1-1 \\
     m_2-m_1-1 \\
   \end{array}\right)\cdots\left(\begin{array}{c}
     i_{\rho(V)}-i_{\rho(V)-1}-1 \\
     m_{\rho(V)}-m_{\rho(V)-1}-1 \\
   \end{array}\right)\left(\begin{array}{c}
     2\nu-i_{\rho(V)} \\
     \nu+1-m_{\rho(V)} \\
   \end{array}\right)\right. - \\
\nonumber & & \hspace{-.7in} \left.
\prod_{j=1}^{\rho(V)}\left(i_{j} -
2m_{j}+1\right)^{|V_{\sigma(j)}|}\left(\begin{array}{c}
     i_1-1 \\
     m_1-1 \\
   \end{array}\right)\left(\begin{array}{c}
     i_2-i_1-1 \\
     m_2-m_1-1 \\
   \end{array}\right)\cdots\left(\begin{array}{c}
     i_{\rho(V)}-i_{\rho(V)-1}-1 \\
     m_{\rho(V)}-m_{\rho(V)-1}-1 \\
   \end{array}\right)\left(\begin{array}{c}
     2\nu-i_{\rho(V)} \\
     \nu+1-m_{\rho(V)} \\
   \end{array}\right)\right)
\end{eqnarray}
where in the last equality we have applied (\ref{beqns5}) and
(\ref{jointdist}).

The formula for $d_V^{(\nu,g)}$ given in Theorem \ref{II.002} now
follows from the next proposition.

\begin{prop}\label{prop4.2}
The inner-most summation may be re-expressed in terms of the following
combinatorial coefficients:
\begin{eqnarray*}
&&\textrm{coeff of}\,\, x^{\nu-r+1} \textrm{in}\\
&&\left(\prod_{n=1}^{r}
\left(i_n-2\sum_{s=1}^n(1+x_s\frac{\partial}{\partial x_s})+ c
\right)^{\alpha_n}\right) \cdot (1+x_1)^{i_1-1} \cdots
(1+x_r)^{i_r-i_{r-1}-1}(1+x_{r+1})^{2\nu-i_r}|_{x_\mu=x}\\
   &=&
\sum_{\begin{array}{c}
    m_1 < \dots < m_{\rho(V)} \\
\in \{1,\dots,\nu+1\}\\
  \end{array}}\\
  &&\prod_{n=1}^{r}\left(i_{n} -
2m_{n}+c\right)^{\alpha_n}\left(\begin{array}{c}
     i_1-1 \\
     m_1-1 \\
   \end{array}\right)\left(\begin{array}{c}
     i_2-i_1-1 \\
     m_2-m_1-1 \\
   \end{array}\right)\cdots\left(\begin{array}{c}
     i_{r}-i_{r-1}-1 \\
     m_{r}-m_{r-1}-1 \\
   \end{array}\right)\left(\begin{array}{c}
     2\nu-i_{r} \\
     \nu+1-m_{r} \\
   \end{array}\right).
\end{eqnarray*}
\end{prop}
To prove this we start with the LHS which, after expanding
binomially, becomes
\begin{eqnarray*}
&&  \textrm{coeff of}\,\, x^{\nu-r+1}
\textrm{in}\,\,\left(\prod_{n=1}^{r}
\left(i_n-2\sum_{s=1}^n(1+x_s\frac{\partial}{\partial x_s})+ c
\right)^{\alpha_n}\right) \cdot \sum_{s_1=0}^{i_1-1}
\sum_{s_2=0}^{i_2-i_1-1} \cdots \sum_{s_r=0}^{i_r-i_{r-1}-1}
\sum_{s_{r+1}=0}^{2\nu-i_r} \\ && \left(\begin{array}{c}
     i_1-1 \\
     s_1 \\
   \end{array}\right)\left(\begin{array}{c}
     i_2-i_1-1 \\
     s_2 \\
   \end{array}\right)\cdots\left(\begin{array}{c}
     i_{r}-i_{r-1}-1 \\
     s_r \\
   \end{array}\right) \left(\begin{array}{c}
     2\nu-i_{r} \\
     s_{r+1} \\
   \end{array}\right) x_1^{s_1}x_2^{s_2}\cdots x_r^{s_r}x_{r+1}^{s_{r+1}}|_{x_\mu=x} \\
   &=& \textrm{coeff of}\,\, x^{\nu-r+1}
\textrm{in}\,\, \sum_{s_1=0}^{i_1-1} \sum_{s_2=0}^{i_2-i_1-1}
\cdots \sum_{s_r=0}^{i_r-i_{r-1}-1} \sum_{s_{r+1}=0}^{2\nu-i_r}
\prod_{n=1}^{r} \left(i_n-2n + c -2\left(s_1 +
\cdots + s_n \right) \right)^{\alpha_n} \cdot\\
&& \left(\begin{array}{c}
     i_1-1 \\
     s_1 \\
   \end{array}\right)\left(\begin{array}{c}
     i_2-i_1-1 \\
     s_2\\
   \end{array}\right)\cdots\left(\begin{array}{c}
     i_{r}-i_{r-1}-1 \\
     s_r\\
   \end{array}\right) \left(\begin{array}{c}
     2\nu-i_{r} \\
     s_{r+1} \\
   \end{array}\right) x_1^{s_1}x_2^{s_2}\cdots x_r^{s_r}x_{r+1}^{s_{r+1}}|_{x_\mu=x} \\
   &=& \textrm{coeff of}\,\, x^{\nu-r+1}
\textrm{in}\,\,\sum_{s_1=0}^{i_1-1} \sum_{s_2=0}^{i_2-i_1-1}
\cdots \sum_{s_r=0}^{i_r-i_{r-1}-1} \sum_{s_{r+1}=0}^{2\nu-i_r}
 \prod_{n=1}^{r} \left(i_n-2n + c -2\left(s_1+
\cdots + s_n \right) \right)^{\alpha_n} \cdot
\\ && \left(\begin{array}{c}
     i_1-1 \\
     s_1 \\
   \end{array}\right)\left(\begin{array}{c}
     i_2-i_1-1 \\
     s_2\\
   \end{array}\right)\cdots\left(\begin{array}{c}
     i_{r}-i_{r-1}-1 \\
     s_r\\
   \end{array}\right) \left(\begin{array}{c}
     2\nu-i_{r} \\
     s_{r+1} \\
   \end{array}\right) x^{s_1 + \cdots + s_{r+1}}\\
   &=& \sum_{s_1 + \cdots +
s_{r+1 = \nu-r+1}}\prod_{n=1}^{r} \left(i_n-2n + c -2\left(s_1+
\cdots + s_n \right) \right)^{\alpha_n} \cdot
\\ && \left(\begin{array}{c}
     i_1-1 \\
     s_1 \\
   \end{array}\right)\left(\begin{array}{c}
     i_2-i_1-1 \\
     s_2\\
   \end{array}\right)\cdots\left(\begin{array}{c}
     i_{r}-i_{r-1}-1 \\
     s_r\\
   \end{array}\right) \left(\begin{array}{c}
     2\nu-i_{r} \\
     s_{r+1} \\
   \end{array}\right)
\end{eqnarray*}
where in the first equality we use the fact that the monomials in
$x_1,\dots, x_{r+1}$ are each simultaneous eigenfunctions for all
the differential operators
$i_n-2\sum_{s=1}^n(1+x_s\frac{\partial}{\partial x_s})+ c$ and
therefore the action of the product of these operators on each
eigenfunction may be replaced by the product of the corresponding
eigenvalues times this function.

We next make the following change of variables in the summations
\begin{eqnarray*}
  s_1 &=& m_1 - 1 \\
  s_2 &=& m_2 - m_1 -1 \\
   &\vdots&  \\
  s_k &=& m_k - m_{k-1} -1 \\
   &\vdots&  \\
  s_r &=& m_r - m_{r-1} -1
\end{eqnarray*}
which then requires that
\begin{eqnarray*}
  s_{r+1} &=& \nu-r+1 - (s_1+\cdots + s_{r}) = \nu-r+1 - (m_r - r)
  = \nu +1 -m_r
\end{eqnarray*}
and this completes the proof of the proposition.

\subsection{Differential Equations for $e_g(t)$ \label{diff_eq_eg}}

We will now show that there are differential equations whose solutions determine
$e_g(-s)$ written in terms of the functions $z_g(s)$. We will show
that these differential equations can be used to derive
expressions for the Taylor Coefficients of $e_g(-s)$,
$\kappa_g^{(\nu)}(j)$, and verify Theorem \ref{II.004thm}.
\medskip

Recall the Hirota relation (\ref{tau_solution}). Restricted to the level
$2\nu$-time flow this becomes
\begin{equation} \label{II.006.1}
b_k(\xi)^2 = \frac{\tau_{k+1}(\xi)^2
\tau_{k-1}(\xi)^2}{\tau_k(\xi)^4} b_k(0)^2
\end{equation}
and is related to the partition
function (\ref{I.001}) through
\begin{equation*}
             \tau_k(\xi)^2 = \frac{Z_k(-2\xi
             k^{\nu-1})}{Z_k(0)}.
\end{equation*}
The logarithm of (\ref{II.006.1}) produces a hierarchy of
difference equations (discrete Hirota relations): Setting $s_k = 2
k^{\nu-1} \xi$ in order to write
\begin{equation} \label{11_2}
\tau_{k}\left(\xi \right)^2 = \frac{Z_{k}( - s_k )}{Z_k(0)},
\end{equation}
and applying a logarithm to both sides of equation (\ref{II.006.1})
one obtains
\begin{equation} \label{eq5}
\log\left(\frac{Z_{k+1}(-s_{k+1})}{Z_{k+1}(0)}\right) - 2
\log\left(\frac{Z_{k}(-s_k)}{Z_{k}(0)} \right) +
\log\left(\frac{Z_{k-1}(-s_{k-1})}{Z_{k-1}(0)}\right)=
\log(b_k(\xi)^2) - \log(b_k(0)^2) .
\end{equation}
\smallskip

We will next describe a continuum limit of these discrete equations which will generate
a hierarchy of differential equations for the functions $e_g(t_{2\nu})$ which appear
in the asymptotic expansion (\ref{I.002}).

If we shift $k \mapsto k \pm 1$, then $s_k$ shifts: $s_k \mapsto
s_{k \pm 1} = s_k (1 \pm 1/k)^{\nu-1}$. With this substitution the left
hand side of (\ref{eq5}) becomes
\begin{equation} \label{step1}
\log\left( \frac{Z_{k+1}(-s(1+1/k)^{\nu-1})}{Z_{k+1}(0)} \right) -
2\log\left(\frac{ Z_k(-s)}{Z_k(0)}\right) + \log\left(
\frac{Z_{k-1}(-s(1-1/k)^{\nu-1})}{Z_{k-1}(0)} \right) .
\end{equation}
Note that from Theorem \ref{I.002}, each term in (\ref{step1})
possesses an asymptotic expansion
of the form
\begin{eqnarray}
& & \log\left( \frac{Z_{k+r}(-s(1+r/k)^{\nu-1})}{Z_{k+r}(0)} \right) =
\sum_{g \ge 0} k^{2 - 2g}\left( 1 + \frac{r}{k} \right)^{2 - 2g}
e_{g}\left( - s \left( 1 + \frac{r}{k}\right)^{\nu  -1} \right),
\end{eqnarray}
with $r = -1, 0, 1$.

The expression (\ref{step1}) is a centered difference. We utilize
a lemma describing centered difference expansions:
\begin{lem} \label{lemma_11_1}
If $\mathcal{G}$ is an analytic function, then
\begin{equation}
\mathcal{G}(1+z) - 2 \mathcal{G}(1) + \mathcal{G}(1-z) = \sum_{n
\geq 1} \frac{2 z^{2n}}{ ( 2n )!} \mathcal{G}^{(2n)}(1)
\end{equation}
as $z \to 0$.
\end{lem}

To apply this lemma we note that (\ref{step1}) has an asymptotic
expansion of the form
\begin{eqnarray}
\nonumber
& & \log\left( \frac{Z_{k+1}(-s(1+1/k)^{\nu-1})}{Z_{k+1}(0)} \right) -
2\log\left(\frac{ Z_k(-s)}{Z_k(0)}\right) + \log\left(
\frac{Z_{k-1}(-s(1-1/k)^{\nu-1})}{Z_{k-1}(0)} \right) \\
\nonumber
& & \hspace{0.4in}
= \sum_{g \ge 0} k^{ 2 - 2g} \left[ \mathcal{G}_{g}\left( 1 - \frac{1}{k}\right)
-2 \mathcal{G}_{g}\left( 1 \right)
+ \mathcal{G}_{g}\left( 1 + \frac{1}{k}\right) \right],
\end{eqnarray}
where
\begin{equation}\label{IV.019}
\mathcal{G}_{g}(w) = w^{2-2g} e_g(-w^{\nu-1} s).
\end{equation}

Using Lemma \ref{lemma_11_1}, we have
\begin{align} \label{eq5L}
\log\left( \frac{Z_{k+1}(-s_k (1+\frac{1}{k})^{\nu-1})}{Z_{k+1}(0)}\right)
- 2 \log\left(\frac{Z_k(-s_k)}{Z_k(0)} \right) + \log\left(
\frac{Z_{k-1}(-s_k(1-\frac{1}{k})^{\nu-1})}{Z_{k-1}(0)} \right) =  \\
\nonumber = \sum_{g \geq 0} \sum_{n \geq 1} k^{2-2g-2n}
\frac{2}{(2n)!} \frac{\partial^{(2n)}}{\partial w^{(2n)}} \left[
w^{2-2n} e_g(-w^{\nu-1} s_k)
  \right] \bigg|_{w=1}.
\end{align}

In terms of the coefficients $z_g$ defined in Proposition \ref{prop_8}, the right hand
side of (\ref{eq5}) becomes
\begin{equation} \label{eq5R}
\log\left( \sum_{g \geq 0} k^{-2g} z_g(s) \right).
\end{equation}

We get ODE's which determine $e_g(-s)$ by equating like orders in
equation (\ref{eq5}) after making the substitutions (\ref{eq5L})
and (\ref{eq5R}).  The equations so derived are
\begin{equation*}
\sum_{n = 0}^g \frac{2}{(2n+2)!} \frac{\partial^{(2n+2)}}{\partial
w^{(2n+2)}} \left[ w^{2-2(g-n)} e_{g-n}(-w^{\nu-1} s) \right]
\bigg|_{w=1} = \mbox{the} \; k^{-2g} \; \mbox{term in} \;
\log\left( \sum_{n = 0} k^{-2n} z_n(s) \right).
\end{equation*}
which establishes the hierarchy (\ref{II.007}).
These equations are
recursive in the sense that we need to know $z_n$ for $n \le g$ and $e_m$
for $m < g$ in order to write a closed differential equation for $e_g$. We will see
how this works out in the next section.

\section{Continuum Toda Solutions and Fine Structure of the Asymptotic Partition Function}

In this section we will prove Theorems \ref{II.003thm} and
\ref{II.004thm}. The equations of the continuum Toda Lattice
hierarchy are given by (\ref{II.006}) and (\ref{Forcing}) and
$b_k(t)$ possesses an asymptotic expansion whose leading order,
$z_0(s)$, is the branch of the solution of the polynomial equation
\begin{equation} \label{V.001}
1 = z_0(s) - c_\nu s z_0(s)^\nu
\end{equation}
which is regular at $s=0$. Recall that
\begin{equation*}
c_\nu = 2\nu \binom{2\nu-1}{\nu-1} .
\end{equation*}
Implicitly differentiating (\ref{V.001}) we have the basic
identity
\begin{equation}\label{V.002B}
ds = \frac{(\nu - (\nu-1)z_0)}{c_\nu z_0^{\nu+1}} dz_0.
\end{equation}
which will be used often in following sections.

\subsection{Integrating the Continuum Toda Lattice Hierarchy}

Let $f_g(s, w) = w^{1-2g} z_g(w^{\nu-1} s)$.  The $k^{-2g}$-order
equation in (\ref{II.006}) is linear in $z_g(s)$:
\begin{equation}\label{V.003}
z_g'(s) = c_\nu \left( f_0^\nu f_{g w} + \nu f_0^{\nu-1} f_g
f_{0w} \right)|_{w=1} + \mbox{\textup{Forcing}}_g|_{w=1}
\end{equation}
where $\mbox{\textup{Forcing}}_g$ is given by (\ref{Forcing}) and
the entire RHS depends only on $z_j, j<g$ and their
derivatives. (Note: in what follows, if the context makes it clear
that the forcing terms are being evaluated at $w=1$ we will simply
write $\mbox{\textup{Forcing}}_g$ rather than
$\mbox{\textup{Forcing}}_g|_{w=1}$.)

Expanding out the derivatives in equation (\ref{V.003}),
\begin{equation}\label{V.004}
z_g' = c_\nu \left( (1-2g) z_0^\nu z_g + (\nu-1) s z_0^\nu z_g' +
\nu z_0^\nu z_g + \nu(\nu-1) s z_0^{\nu-1} z_g z_0' \right) +
\mbox{\textup{Forcing}}_g.
\end{equation}
Solving for $z_g'$:
\begin{equation} \label{V.004.1}
(1 - c_\nu (\nu-1) s z_0^\nu ) z_g' = c_\nu \left( (\nu + 1 - 2g)
z_0^\nu +
  \nu (\nu-1) s z_0^{\nu-1} z_0' \right) z_g + \mbox{\textup{Forcing}}_g.
\end{equation}

We may inductively assume that $z_j(s)$ is a function of $z_0(s)$
for $j<g$, and as a result $\mbox{Forcing}_g$ is a function of
$z_0$. One can convert equation (\ref{V.004.1}) to a
differential equation for $z_g$ as a function of $z_0$ using
equations (\ref{V.001}) and (\ref{V.002B}):

\begin{equation}
\label{V.004.2} \frac{dz_g}{dz_0}= \frac{ \left( \nu (2-2g) +
(\nu-1)(2g-1)z_0\right)}{z_0
  (\nu-(\nu-1) z_0)} z_g + \frac{ \mbox{\textup{Forcing}}_g}{c_\nu
  z_0^{\nu+1}}.
\end{equation}

Equation (\ref{V.004.2}) is linear in $z_g$. Let
\begin{align} \nonumber
G(z_0, y) &= \int_{y}^{z_0} \frac{ \left( \nu (2-2g) +
  (\nu-1)(2g-1)u\right)}{u (\nu-(\nu-1) u)} du
\\
&= \int_{y}^{z_0} \left( (\nu-1)^{-1} (\nu-(\nu-1)u)^{-1} + 2(1-g)
u^{-1} \right) du
\\ \label{V.005}
&= - \log(\nu-(\nu-1)u) + 2(1-g) \log(u) \bigg|_{u=y}^{u=z_0}.
\end{align}
Integrating equation (\ref{V.004}) one has
\begin{align} \nonumber
z_g(s) &= \int_1^{z_0} \exp( G(z_0, y) ) \frac{1}{c_\nu y^{\nu+1}
} \mbox{\textup{Forcing}}_g(y) dy
\\
&= \frac{z_0^{2(1-g)} }{\nu-(\nu-1)z_0} \int_1^{z_0}
\frac{\nu-(\nu-1)
  y}{c_\nu y^{\nu+3-2g}} \mbox{\textup{Forcing}}_g(y) dy,
\label{zg}
\end{align}
where $z_0 = z_0(s)$. This establishes the basic formula stated in
Theorem \ref{II.003thm}(3).

\subsection{Singularities of $z_g$}

In this section we complete the proof of Theorem \ref{II.003thm}.
Expression (\ref{zg}) can be exploited to determine the
singularity structure of $z_g(s)$. We differentiate equation
(\ref{V.001}) and solve for $z_0'(s)$:
\begin{equation} \label{V.005.1}
z_0'(s) = \frac{ c_\nu z_0^{\nu+1}}{(\nu- (\nu-1) z_0)}.
\end{equation}
With equation (\ref{V.005.1}) we can re-express successive
derivatives of $z_0$ as functions of just $z_0$ by using the chain
rule.

Doing so, we find the following result:
\begin{lem} \label{V.lem1}
The derivatives of $z_0(s)$ have the following form
\begin{equation} \label{V.005.2}
z_0^{(j)}(s) = \frac{ c_\nu^j z_0^{\nu j + 1} }{(\nu - (\nu-1)
z_0)^{2 j - 1} } \cdot \left(\mbox{ Polynomial of degree }\; j-1
\right).
\end{equation}
Equation (\ref{V.005.2}) together with (\ref{V.002B}) give the
following result
\begin{equation} \label{V.005.3}
s^j z_0^{(j)}(s) = \frac{ z_0 (z_0 - 1)^j }{(\nu - (\nu-1)
z_0)^{2j-1}} \cdot
  \left( \mbox{ Polynomial of degree }\; j-1 \right).
\end{equation}
\end{lem}
\medskip

The terms in $\mbox{\textup{Forcing}}_g$ (\ref{Forcing}) are each of the form
\begin{equation} \label{V.005.4.A}
(f_{k_1})_{w^{(n_1)}} (f_{k_2})_{w^{(n_2)}} \dots
  (f_{k_{\nu+1}})_{w^{(n_{\nu+1})}} \bigg|_{w=1},
\end{equation}
where $f_k=w^{1-2k} z_k(w^{\nu-1} s)$, and $n_1 + n_2 + \dots +
n_{\nu+1} = 2(g-k_1 - k_2 - \dots - k_{\nu+1} )$. Expanding the
derivatives in (\ref{V.005.4.A}) one finds that
$\mbox{\textup{Forcing}}_g$ is in fact comprised of terms of the form:
\begin{equation} \label{V.005.4}
s^j z_{k_1}^{(j_1)} z_{k_2}^{(j_2)} \dots
z_{k_{\nu+1}}^{(j_{\nu+1})},
\end{equation}
where
\begin{align*}
j &= j_1 + j_2 + \dots + j_{\nu+1}, \\
0 \leq j &\leq 2(g - k_1 - k_2 - \dots - k_{\nu+1}).
\end{align*}

\begin{lem}\label{V.prop1}
The function $z_g^{(j)}(s)$ ($j$ derivatives of $z_g(s)$) is of
class \emph{iir} with singularities occurring only at $z_0 = 0$ or
$z_0=\nu/(\nu-1)$.
\end{lem}

We will first prove by induction that $z_g(s)$ satisfies
Lemma \ref{V.prop1}; see section \ref{g=1} for the initial
step. Assume that $z_k^{(j)}(s)$ for $k<g$ satisfies Lemma
\ref{V.prop1}; this assumption, Lemma \ref{V.lem1} and formula
(\ref{V.001}) show that the integrand in formula (\ref{zg}) is a
polynomial in functions of class \emph{iir} with singularities
restricted to $0$ and $\nu/(\nu-1)$ and having rational
coefficients with at worst poles located at $z_0 = 0$. Thus the
integral remains of class \emph{iir} with the stated singularities
and (\ref{zg}) is then this integral multiplied by a rational
function with pole only at $\nu/(\nu-1)$.

We next prove, by induction on $j$, that the higher derivatives,
$z_g^{(j)}(s)$, satisfy Lemma \ref{V.prop1} which will
complete the original induction. A derivative with respect to
$z_0$ will raise the order of a singularity but will not introduce
new singularities. Compute
\begin{equation*}
z_g^{(j)}(s) = \frac{d}{dz_0}\left[ z_g^{(j-1)}(s) \right]
z_0'(s);
\end{equation*}
by the induction step (in $j$) and formula (\ref{V.005.2}) this
expression is of class \emph{iir} with singularities only at $z_0
= 0$ and $z_0 = \nu/(\nu-1)$.

With the forthcoming verification of the initial step of the
induction (subsection \ref{g=1}), this completes the proof of Theorem
\ref{II.003thm}.

\subsection{Example: $g=1$ \label{g=1}}

Recall that $f_0(s, w) = w z_0(w^{\nu-1} s)$ and $f_1(s, w) =
w^{-1} z_1(w^{\nu-1} s)$. The forcing for $z_1$ is given
exclusively in terms of $z_0$ as
\begin{equation} \label{Forcing1}
c_\nu^{-1} \mbox{\textup{Forcing}}_1= c_\nu^{-1} F_1^\nu[0] =
\frac{\nu}{6} f_0^\nu {f_0}_{www} + \frac{\nu(\nu-1)}{3}
f_0^{\nu-1} {f_0}_w {f_0}_{ww} + \frac{\nu(\nu-1)(\nu-2)}{12}
f_0^{\nu-2} {f_0}_w^3.
\end{equation}
The entire expression is evaluated at $w=1$. We find that
\begin{align} \label{Forcing1_expanded}
c_\nu^{-1} F_1^\nu[0] \big|_{w=1} &= \frac{\nu(\nu-1)(\nu-2)}{12} z_0^{\nu+1}
+ \frac{\nu
  (\nu-1)}{12} (9 \nu^2 - 17 \nu + 6 ) s z_0^\nu z_0' + \frac{5 \nu
  (\nu-1)^3}{6} s^2 z_0^\nu z_0''
\\ & \phantom{=} \nonumber
+ \frac{\nu
  (\nu-1)^3}{12} (7\nu -6) s^2 z_0^{\nu-1} (z_0')^2 + \frac{\nu (\nu-1)^3}{6}
  s^3 z_0^\nu z_0''' + \frac{\nu (\nu-1)^4}{3} s^3 z_0^{\nu-1} z_0' z_0''
\\ & \phantom{=} \nonumber
 +
  \frac{\nu (\nu-1)^4 (\nu-2)}{12} s^3 z_0^{\nu-2} (z_0')^3.
\end{align}

Inserting (\ref{Forcing1_expanded}) into equation (\ref{zg}) with
$g=1$, (writing all derivatives out as rational expressions of
$z_0$) yields
\begin{equation} \label{z1}
z_1(t) = \frac{ (\nu-1) \nu (z_0 -1) z_0 (-\nu^2 - 2z_0 +\nu z_0 +
\nu^2 z_0 )}{12 (\nu - (\nu-1)z_0 )^4}.
\end{equation}

>From this we see that not only does $z_1(t)$ satisfy Lemma
\ref{V.prop1}, but in fact this expression is a rational function,
without any of the potential complexities of the   \emph{iir}
class. In fact all the examples we have worked out here for low
values of $g$ yield rational expressions in $z_0$ for $z_g$.
Moreover, the poles of these expressions are restricted to
$z_0=\nu/(\nu -1)$.
\bigskip

{\bf Remark} The rationality of (\ref{z1}) seems to depend very
sensitively on the precise coefficients appearing in
$\mbox{Forcing}_g$. We illustrate this by looking at the
contributions that particular terms of (\ref{Forcing1_expanded})
give to (\ref{z1}). The integrand in integral (\ref{zg}) has a
term of the form
\begin{equation*}
\frac{(\nu - (\nu-1) z_0)}{z_0^{\nu+1}} s^2 z_0^{\nu-1} (z_0')^2 =
\frac{(z_0-1)^2 } {(\nu-(\nu-1) z_0)} ,
\end{equation*}
where we have used equation (\ref{V.005.1}) to eliminate $z_0'$.
This term produces a logarithmic singularity at $\nu/(\nu -1)$;
therefore one of the other terms must cancel this one to eliminate
the logarithm. $\Box$
\medskip

>From the representation (\ref{z1}) we can compute the Taylor
Coefficients of $z_1(t)$ using the technique outlined in formulas
(\ref{dreln}-\ref{zcoeff}):
\begin{align} \nonumber
\zeta^{(1)}_j &= \frac{c_\nu^j}{2\pi i} \oint_{u \sim 1}
\frac{(\nu-(\nu-1)
  u) u^{\nu j - 1}}{(u-1)^{j+1}}  z_1(z_0=u) du \\ \nonumber
&= \frac{c_\nu^j}{2\pi i} \oint_{u \sim 1} \frac{ (\nu-1) \nu
u^{\nu j}
  (-\nu^2 - 2u + \nu u + \nu^2 u )}{ 12 (u-1)^j (\nu - (\nu-1) u)^3} du  \\
\label{tcz1} &= \frac{c_\nu^j}{2\pi i} \oint_{w \sim 0} \frac{
(\nu-1) \nu (w+1)^{\nu j} ( \nu-2
  + (\nu^2 + \nu -2) w)}{12 (1-(\nu-1) w)^3 w^j } dw.
\end{align}
In the last line we have made the change of variables $w = u-1$.
With the integral representation (\ref{tcz1}) one may conclude
that
\begin{align*}
\zeta^{(1)}_j = \nu c_\nu^j\left[ w^{j-1} \mbox{Taylor Coefficient
of } \frac{ \left(  (\nu-1) (\nu-2) + (\nu+2)w  \right)  (w+1)^{\nu j} }
{(1-(\nu-1) w)^3}\right].
\end{align*}

\subsection{Example:  $g=2$}

The forcing terms for $z_2$  (\ref{Forcing}) are:
\begin{enumerate}
\item The first term is the summation in (\ref{Forcing}):
\begin{equation*}
c_\nu \nu f_0^{\nu-1} f_1 {f_1}_w + c_\nu \nu (\nu-1)/2 f_0^{\nu-2} {f_0}_w
f_1^2.
\end{equation*}

\item The middle term is $F_1^{(\nu)}[2]$:
\begin{align*}
c_\nu^{-1} F_1^{(\nu)}[2] &= \frac{\nu^2}{6} f_0^{\nu-1} f_1 {f_{0}}_{www} +
\frac{\nu}{6} f_0^\nu {f_1}_{www} + \frac{\nu(\nu-1)^2}{3} f_0^{\nu-2} f_1
     {f_0}_w {f_0}_{ww} + \frac{\nu (\nu-1)}{3} f_0^{\nu-1} {f_1}_w {f_0}_{ww}
     \\ &\phantom{=}
+ \frac{\nu (\nu-1)}{3} f_0^{\nu-1} {f_0}_w {f_1}_{ww} +
     \frac{\nu(\nu-1)(\nu-2)^2 }{12} f_0^{\nu-3} f_1 {f_0}_w^3 +
     \frac{\nu(\nu-1)(\nu-2)}{4} f_0^{\nu-2} {f_0}_w^2 {f_1}_w.
\end{align*}

\item The last term arises from $F_2^{(\nu)}[0]$:
\begin{align*}
\frac{1}{c_\nu} F_2^{(\nu)}[0] \big|_{w=1} =& \frac{\nu (2\nu-1)}{120} f_0^\nu
     {f_0}_{w^{(5)}} +
\frac{\nu(\nu-1)(18\nu-11)}{360} f_0^{\nu-1} {f_0}_w
{f_0}_{w^{(4)}} + \frac{\nu(\nu-1)(6\nu-5)}{72} f_0^{\nu-1}
{f_0}_{ww} {f_0}_{www}
\\ &
+ \frac{\nu(\nu-1)(\nu-2)(23\nu-20)}{360} f_0^{\nu-2} {f_0}_w^2
{f_0}_{www} + \frac{\nu(\nu-1)(\nu-2)(62\nu-65)}{720} f_0^{\nu-2}
{f_0}_w {f_0}_{ww}^2
\\ &
+ \frac{\nu(\nu-1)(\nu-2)(\nu-3)(16\nu-19)}{360} f_0^{\nu-3}
{f_0}_{ww} {f_0}_w^3
\\ &
 +
\frac{\nu(\nu-1)(\nu-2)(\nu-3)(\nu-4)(5\nu-7)}{1440} f_0^{\nu-4}
{f_0}_w^5.
\end{align*}

\end{enumerate}

The three terms make up the forcing terms for $z_2$; each of them
is expressed as a function of $z_0$. Substituting this explicit
representation into  equation (\ref{zg}) with $g=2$ we find that
\begin{align*}
z_2(t) =& \frac{1}{1440} (\nu-1) \nu (z_0-1) z_0 \left[ (2 \nu^6 -
14 \nu^7 + 24 \nu^8)
\right.  \\
& \left. +(-12 \nu^3 + 148 \nu^4 -546 \nu^5 + 758 \nu^6 - 252
\nu^7 - 96 \nu^8 ) z_0
\right. \\
& \left. + (264 \nu^2 - 1510 \nu^3 + 25551 \nu^4 - 500 \nu^5 -1789
\nu^6 + 840 \nu^7 + 144 \nu^8 ) z_0^2
\right. \\
& \left. + (-536 \nu + 1396 \nu^2 + 912 \nu^3 -4596 \nu^4 + 2492
\nu^5 + 1296 \nu^6 - 868 \nu^7 - 96 \nu^8 ) z_0^3
\right. \\
& \left. + (168 + 234 \nu - 1467 \nu^2 + 558 \nu^3 + 1902 \nu^4 -
1446 \nu^5 - 267 \nu^6 + 294 \nu^7 + 24 \nu^8 ) z_0^4 \right]
 \\
& \cdot (\nu-(\nu-1)z_0)^{-9}.
\end{align*}

Let
\begin{align*}
r_2(w) &= \frac{c_\nu^j \nu (\nu-1)}{1440}
(w+1)^{\nu j}  \bigg[
(\nu-2)(\nu-3)(\nu-4) (5\nu -7)
\\ & \phantom{=}
+ 2(\nu-1)(\nu-2) ( 168 + 84\nu - 246 \nu^2 + 73 \nu^3 )
w
\\ & \phantom{=}
+ (\nu-1)^2 (\nu-2)(-504 - 1158 \nu + 288 \nu^2 + 497 \nu^3)
w^2
\\ & \phantom{=}
+ 4(\nu-1)^3 ( - 168 - 604 \nu - 190 \nu^2 + 288 \nu^3 + 77 \nu^4 )
w^3
\\ & \phantom{=}
+ 3(\nu-1)^4 ( 56 + 302 \nu + 383 \nu^2 + 130 \nu^3 + 8 \nu^4 ) w^4
 \bigg] (1 - (\nu-1) w)^{-8} .
\end{align*}
We compute the Taylor Coefficients of $z_2(t)$ with a loop
integral:
\begin{align*}
\zeta_j^{(2)} &= \frac{c_\nu^j}{2\pi i} \oint_{u \sim 1} \frac{(\nu
- (\nu-1) u)
  u^{\nu j -1} }{ (u-1)^{j+1} } z_2(z_0= u) du
\\ &= \mbox{The $w^{j-1}$ coefficient of $r_2(w)$}.
\end{align*}

\subsection{Example: $g=3$}

To compute $z_3$ we compute the forcing as above.  Here we outline
how the terms contributing to this forcing are found and then present the
resulting explicit expression for $z_3$.  There are
six terms in formula (\ref{Forcing}) when $g=3$:

\begin{enumerate}

\item[(1-2)]
The first two terms occur in the summation in (\ref{Forcing}) when
$g=3$.  One term contains three $f_1$'s and is
\begin{align*}
& \frac{\nu (\nu-1) (\nu-2) }{6} f_0^{\nu-3} f_1^3 {f_0}_w + \frac{\nu
    (\nu-1)}{2} f_0^{\nu-2} f_1^2 {f_1}_w.
\end{align*}
The other, containing an $f_1$ and an $f_2$, is
\begin{align*}
&\nu (\nu-1) {f_0}^{\nu-2} f_1 f_2 {f_0}_w + \nu f_0^{\nu-1} f_1 {f_2}_w + \nu
f_0^{\nu-1} f_2 {f_1}_w .
\end{align*}

\item[(3-4)] The next two terms are contributed by $F_1^{(\nu)}[4]$:
this expression represents the fourth order contributions of $F_1^{(\nu)}$.
The terms involving two $f_1$'s are
\begin{align*}
& \frac{\nu^2}{6} f_0^{\nu-1} f_1 {f_1}_{www}
+ \frac{\nu^2(\nu-1)}{12}  f_0^{\nu-2} f_1^2 {f_0}_{www}
+ \frac{\nu (\nu-1)}{3} f_0^{\nu-1} {f_1}_w {f_1}_{ww}
+ \frac{\nu (\nu-1)^2}{3} f_0^{\nu-2} f_1 {f_0}_w {f_1}_{ww}
\\&+ \frac{\nu (\nu-1)^2}{3} f_0^{\nu-2} f_1 {f_1}_w {f_0}_{ww}
+ \frac{\nu (\nu-1)^2 (\nu-2)}{6} f_0^{\nu-3} f_1^2 {f_0}_w {f_0}_{ww}
+ \frac{\nu (\nu-1) (\nu-2)}{4} f_0^{\nu-2} {f_0}_w {f_1}_w^2
\\&+ \frac{\nu (\nu-1) (\nu-2)^2}{4} f_0^{\nu-3} f_1 {f_0}_w^2 {f_1}_w
+ \frac{\nu (\nu-1) (\nu-2)^2 (\nu-3)}{24} f_0^{\nu-4} f_1^2 {f_0}_w^3.
\end{align*}
The terms containing one $f_2$ are
\begin{align*}
& \frac{\nu}{6} f_0^\nu {f_2}_{www}
+ \frac{\nu^2}{6} f_0^{\nu-1} f_2 {f_0}_{www}
+ \frac{\nu(\nu-1)}{3} f_0^{\nu-1} {f_0}_w {f_2}_{ww}
+ \frac{\nu (\nu-1)}{3} f_0^{\nu-1} {f_2}_w {f_0}_{ww}
\\&+ \frac{\nu (\nu-1)^2}{3} f_0^{\nu-2} f_2 {f_0}_w {f_0}_{ww}
+ \frac{\nu (\nu-1)(\nu-2) }{4} f_0^{\nu-2} {f_0}_w^2 {f_2}_w
+ \frac{\nu (\nu-1)(\nu-2)^2}{12} f_0^{\nu-3} f_2 {f_0}_w^3.
\end{align*}

\item[(5)]  The fifth term is
\begin{align*}
F_2^{(\nu)}[2] &=
\frac{\nu (2\nu-1) }{120} f_0^\nu {f_1}_{w^{(5)}}
+ \frac{\nu^2 (2\nu-1)}{120} f_0^{\nu-1} f_1 {f_0}_{w^{(5)}}
+ \frac{\nu (\nu-1) (18\nu-11)}{360} f_0^{\nu-1} {f_0}_w {f_1}_{w^{(4)}}
\\ &\phantom{=}
+ \frac{\nu (\nu-1) (18\nu -11)}{360} f_0^{\nu-1} {f_1}_w {f_0}_{w^{(4)}}
+ \frac{\nu (\nu-1)^2 (18\nu -11)}{360} f_0^{\nu-2} f_1 {f_0}_w
{f_0}_{w^{(4)}}
\\ &\phantom{=}
+ \frac{\nu (\nu-1) (6\nu -5)}{72} f_0^{\nu-1} {f_0}_{ww} {f_1}_{www}
+ \frac{\nu (\nu-1) (6\nu -5)}{72} f_0^{\nu-1} {f_1}_{ww} {f_0}_{www}
\\ &\phantom{=}
+ \frac{\nu (\nu-1)^2 (6\nu-5)}{72} f_0^{\nu-2} f_1 {f_0}_{ww} {f_0}_{www}
+ \frac{\nu (\nu-1)(\nu-2)(23\nu -20)}{360} f_0^{\nu-2} {f_0}_w^2 {f_1}_{www}
\\ &\phantom{=}
+ \frac{\nu (\nu-1)(\nu-2)(23\nu -20)}{180} f_0^{\nu-2} {f_0}_w {f_1}_w
{f_0}_{www}
+ \frac{\nu (\nu-1)(\nu-2)^2(23\nu -20)}{360} f_0^{\nu-3} f_1 {f_0}_w^2
{f_0}_{www}
\\ &\phantom{=}
+ \frac{\nu (\nu-1)(\nu-2)(62\nu -65)}{360} f_0^{\nu-2} {f_0}_w {f_0}_{ww}
{f_1}_{ww}
+ \frac{\nu (\nu-1)(\nu-2)(62\nu -65)}{720} f_0^{\nu-2} {f_1}_w {f_0}_{ww}^2
\\ &\phantom{=}
+ \frac{\nu (\nu-1)(\nu-2)^2(62\nu-65)}{720} f_0^{\nu-3} f_1 {f_0}_w
{f_0}_{ww}^2
+ \frac{\nu (\nu-1)(\nu-2)(\nu-3) (16\nu -19)}{360} f_0^{\nu-3} {f_0}_w^3
{f_1}_{ww}
\\ &\phantom{=}
+ \frac{\nu (\nu-1)(\nu-2)(\nu-3) (16\nu -19)}{120} f_0^{\nu-3} {f_0}_w^2
{f_1}_w {f_0}_{ww}
\\ &\phantom{=}
+ \frac{\nu(\nu-1)(\nu-2)(\nu-3)^2(16\nu -19)}{360} f_0^{\nu-4} f_1 {f_0}_w^3
{f_0}_{ww}
\\ &\phantom{=}
+ \frac{\nu (\nu-1)(\nu-2)(\nu-4)(5\nu-7)}{288} f_0^{\nu-4} {f_0}_w^4 {f_1}_w
+ \frac{\nu (\nu-1)(\nu-2)(\nu-4)^2 (5\nu -7)}{1440} f_0^{\nu-5} f_1 {f_0}_w^5.
\end{align*}

\item[(6)]   The last term is
\begin{align*}
F_3^{(\nu)}[0] &=
\frac{\nu}{7!} (3-8\nu + 6\nu^2) f_0^\nu {f_0}_{w^{(7)}}
+ \frac{\nu (\nu-1)}{7! 3} (47-108\nu +72\nu^2) f_0^{\nu-1} {f_0}_w {f_0}_{w^{(6)}}
\\ &\phantom{=}
+ \frac{\nu (\nu-1)}{7! 2} (105 -210\nu + 112\nu^2 ) f_0^{\nu-1}
{f_0}_{ww} {f_0}_{w^{(5)}}
+ \frac{\nu (\nu-1)}{7! 6} (539 -1050 \nu + 504 \nu^2) f_0^{\nu-1}
{f_0}_{w^{(3)}} {f_0}_{w^{(4)}}
\\ &\phantom{=}
+ \frac{\nu (\nu-1)(\nu-2)}{7! 2} (86 -163\nu + 86 \nu^2 ) f_0^{\nu-2}
{f_0}_w^2 {f_0}_{w^{(5)}}
\\ &\phantom{=}
+ \frac{\nu (\nu-1)(\nu-2)}{7! 6} (1387 - 2416\nu +1044\nu^2) f_0^{\nu-2}
{f_0}_w {f_0}_{ww} {f_0}_{w^{(4)}}
\\ &\phantom{=}
+ \frac{\nu (\nu-1)(\nu-2)}{7! 3} (467 - 803\nu + 327 \nu^2 ) f_0^{\nu-2}
{f_0}_w {f_0}_{w^{(3)}}^2
\\ &\phantom{=}
+ \frac{\nu (\nu-1)(\nu-2)}{7! 6} (1456 - 2359 \nu + 882 \nu^2) f_0^{\nu-2}
{f_0}_{ww}^2 {f_0}_{w^{(3)}}
\\ &\phantom{=}
+ \frac{\nu (\nu-1)(\nu-2)(\nu-3)}{7! 6} (410 -665 \nu + 270 \nu^2 )
f_0^{\nu-3} {f_0}_w^3 {f_0}_{w^{(4)}}
\\ &\phantom{=}
+ \frac{\nu (\nu-1)(\nu-2)(\nu-3)}{7! 2} (857 - 1309 \nu + 458 \nu^2 )
f_0^{\nu-3} {f_0}_w^2 {f_0}_{ww} {f_0}_{w^{(3)}}
\\ &\phantom{=}
+ \frac{\nu (\nu-1)(\nu-2)(\nu-3)}{7! 6} (1327 - 1932 \nu + 620 \nu^2 )
f_0^{\nu-3} {f_0}_w {f_0}_{ww}^3
\\ &\phantom{=}
+ \frac{\nu (\nu-1)(\nu-2)(\nu-3)(\nu-4)}{7! 12} (788 -1121 \nu + 359 \nu^2 )
f_0^{\nu-4} {f_0}_w^4 {f_0}_{w^{(3)}}
\\ &\phantom{=}
+ \frac{\nu (\nu-1)(\nu-2)(\nu-3)(\nu-4)}{7! 12} (2431 -3315 \nu + 974 \nu^2)
f_0^{\nu-4} {f_0}_w^3 {f_0}_{ww}^2
\\ &\phantom{=}
+ \frac{\nu (\nu-1)(\nu-2)(\nu-3)(\nu-4)(\nu-5)}{7! 6} (228 -290 \nu + 77
\nu^2 ) f_0^{\nu-5} {f_0}_w^5 {f_0}_{ww}
\\ &\phantom{=}
+ \frac{\nu (\nu-1)(\nu-2)(\nu-3)(\nu-4)(\nu-5)(\nu-6)}{7! 72} (124 -147 \nu +
35 \nu^2 ) f_0^{\nu-6} {f_0}_w^7.
\end{align*}

\end{enumerate}

The entire forcing expression is a rational function
of $z_0$ with singularities only at $z_0 = \nu/(\nu-1)$.  We
insert these forcing terms into equation (\ref{zg}) with $g=3$
and find that:
\begin{align*}
z_3(t) &= \frac{\nu (\nu-1)}{362880} \frac{ z_0 (z_0
  -1)}{(\nu-(\nu-1)z_0)^{14}}
\\ &\phantom{=} \left[
(\nu-2)(\nu-3)(\nu-4)(\nu-5)(\nu-6)(124-147\nu + 35\nu^2)
\right. \\ & \phantom{=[] }
+ (\nu-3)(\nu-2)(\nu-1) (104160 +47584\nu -332550\nu^2 +
270697\nu^3 - 83226\nu^4 + 8923 \nu^5 ) (z_0 - 1)
\\ & \phantom{=[]}
+ 3(\nu-2)(\nu-1)^2(312480+744980\nu -1245750\nu^2 + 373091\nu^3 +
1085920\nu^4 - 485414 \nu^5
\\ & \phantom{=[]+ 3(\nu-2)(\nu-1)^2(312480)}
+ 67225 \nu^6 )(z_0-1)^2
\\ & \phantom{=[]}
+ (\nu-2)(\nu-1)^3(-1562400 - 7251840\nu +290690\nu^2 + 11468057
\nu^3 - 2824078\nu^4 -3154302\nu^5
\\ & \phantom{=[]+ (\nu-2)(\nu-1)^3(-1562400)}
+ 1078663 \nu^6 )(z_0-1)^3
\\ & \phantom{=[]}
+ (\nu-2)(\nu-1)^4 (1562400 + 10781280 \nu + 12588010\nu^2 -
10677353 \nu^3 - 11255921 \nu^4 + 3006363 \nu^5
\\ & \phantom{=[]+ (\nu-2)(\nu-1)^4 (1562400)}
+ 1779986 \nu^6 )(z_0-1)^4
\\ &\phantom{=[]}
+ 3(\nu-1)^5 (624960 + 5411808 \nu + 10100796 \nu^2 - 1315908
\nu^3 - 9371695 \nu^4 - 973573 \nu^5 + 1835799 \nu^6
\\ &\phantom{=[] + 3(\nu-1)^5 (624960)}
+ 308858 \nu^7 )(z_0-1)^5
\\ &\phantom{=[]}
+ (\nu-1)^6 (-624960 - 6823584\nu -20098900 \nu^2 - 16851720 \nu^3
+ 3867117 \nu^4 + 8356442 \nu^5
\\ &\phantom{=[]+ (\nu-1)^6 (-624960)}
+ 2223760 \nu^6 + 119824 \nu^7 )(z_0 - 1)^6
\\ &\phantom{=[]}
 + 5(\nu-1)^7 (17856 + 235296 \nu + 939236 \nu^2 + 1505064 \nu^3 + 1032603
\nu^4 + 285860 \nu^5 + 24472 \nu^6
\\ &\phantom{=[]+ 5(\nu-1)^7 (17856)} \left.
+ 64 \nu^7 )(z_0 -1)^7 \right].
\end{align*}


Let
\begin{align*}
r_3(w) &=  \frac{c_\nu^j \nu (\nu-1)}{ 362880}
(w+1)^{\nu j}
  \bigg[
(\nu-2)(\nu-3)(\nu-4)(\nu-5)(\nu-6)(124-147\nu + 35\nu^2)
\\ & \phantom{=}
+ (\nu-3)(\nu-2)(\nu-1) (104160 +47584\nu -332550\nu^2 +
270697\nu^3 - 83226\nu^4 + 8923 \nu^5 )
w
\\ & \phantom{=}
+ 3(\nu-2)(\nu-1)^2(312480+744980\nu -1245750\nu^2 + 373091\nu^3 +
1085920\nu^4 - 485414 \nu^5
+ 67225 \nu^6 )
w^2
\\ & \phantom{=}
+ (\nu-2)(\nu-1)^3(-1562400 - 7251840\nu +290690\nu^2 + 11468057
\nu^3 - 2824078\nu^4 -3154302\nu^5
\\ & \phantom{=[]+ (\nu-2)(\nu-1)^3(-1562400)}
+ 1078663 \nu^6 )
w^3
\\ & \phantom{=}
+ (\nu-2)(\nu-1)^4 (1562400 + 10781280 \nu + 12588010\nu^2 -
10677353 \nu^3 - 11255921 \nu^4 + 3006363 \nu^5
\\ & \phantom{=[]+ (\nu-2)(\nu-1)^4 (1562400)}
+ 1779986 \nu^6 )
w^4
\\ & \phantom{=}
+ 3(\nu-1)^5 (624960 + 5411808 \nu + 10100796 \nu^2 - 1315908
\nu^3 - 9371695 \nu^4 - 973573 \nu^5 + 1835799 \nu^6
\\ &\phantom{=[] + 3(\nu-1)^5 (624960)}
+ 308858 \nu^7 )
w^5
\\ & \phantom{=}
+ (\nu-1)^6 (-624960 - 6823584\nu -20098900 \nu^2 - 16851720 \nu^3
+ 3867117 \nu^4 + 8356442 \nu^5
\\ &\phantom{=[]+ (\nu-1)^6 (-624960)}
+ 2223760 \nu^6 + 119824 \nu^7 )
w^6
\\ & \phantom{=}
+ 5(\nu-1)^7 (17856 + 235296 \nu + 939236 \nu^2 + 1505064 \nu^3 + 1032603
\nu^4 + 285860 \nu^5 + 24472 \nu^6
\\ &\phantom{=[]+ 5(\nu-1)^7 (17856)}
+ 64 \nu^7 )
w^7
\bigg]  (1 - (\nu-1) w)^{-13}.
\end{align*}
We compute the Taylor
Coefficients of $z_3(t)$ with a loop integral:
\begin{align*}
\zeta^{(3)}_j &= \frac{c_\nu^j}{2\pi i} \oint_{u \sim 1} \frac{(\nu -
  (\nu-1) u) u^{\nu j - 1}}{ (u-1)^{j+1}} z_3(z_0 = u) du \\
& = \mbox{The $w^{j-1}$ coefficient of $r_3(w)$}.
\end{align*}

\subsection{Determining $e_g(-s_{2\nu})$}

Theorem \ref{II.003Bthm} gives a second order differential
equation determining $e_g(-s_{2\nu})$:
\begin{equation}\label{eg_equation}
\frac{\partial^2}{\partial w^2} \left[ w^{2-2g}
e_g\left(-w^{\nu-1} s\right)
  \right] = \mbox{\textup{drivers}}_g,
\end{equation}
where the entire equation is evaluated at $w=1$ and where the
drivers are the right hand side of equation (\ref{II.007}):
\begin{equation*}
- \sum_{n=1}^g \frac{2}{(2n+2)!} \frac{\partial^{2n+2}}{\partial
w^{2n+2}}
  \left[ w^{2-2(g-n)} e_{g-n}(-w^{\nu-1} s) \right] + \mbox{the $k^{-2g}$ term
  of}\quad \log\left[ \sum_{n=0}^\infty \frac{1}{k^{2n}} z_n(s) \right].
\end{equation*}

Expanding the two derivatives on the left hand side of
(\ref{eg_equation})
 and setting $w=1$ we arrive at the basic ode for the $e_g$:
\begin{equation}\label{V.006}
(2-2g)(1-2g) e_g(-s) - (\nu-1) (\nu+2-4g) s e_g'(-s) + (\nu-1)^2
s^2
  e_g''(-s)  =
  \mbox{\textup{drivers}}_g.
\end{equation}

If $g\neq 1$ one solves this equation by integrating factors as
follows: Multiply both sides of (\ref{V.006}) by $s^{\gamma_1}$
with
\begin{equation} \label{gamma_1}
\gamma_1 = \gamma_1^{(+)} =\frac{3-\nu -2g}{\nu-1} \quad \mbox{or}
\quad  \gamma_1^{(-)} = \frac{2-\nu
-
  2g}{\nu-1},
\end{equation}
to arrive at
\begin{equation}\label{V.006B}
(2-2g)(1-2g) s^{\gamma_1} e_g(-s) - (\nu-1)(\nu+2-4g) s^{\gamma_1+1}
e_g'(-s)
  + (\nu-1)^2
  s^{\gamma_1+2} e_g''(-s) = s^{\gamma_1} \mbox{\textup{drivers}}_g.
\end{equation}
Let
\begin{equation*}
A = (2-2g)(1-2g)/(\gamma_1+1).
\end{equation*}
We then
integrate equation (\ref{V.006B}) once,
\begin{equation} \label{V.007}
A s^{\gamma_1+1} e_g(-s) - (\nu-1)^2 s^{\gamma_1+2} e_g'(-s) =
\int_0^s {s_1}^{\gamma_1} \mbox{\textup{drivers}}_g(s_1) ds_1 + K_1',
\end{equation}
where $K_1'$ is a constant of integration. Multiplying by
$s^{\gamma_2}$,
\begin{equation}\label{V.008}
A s^{\gamma_1+\gamma_2+1} e_g(s) - (\nu-1)^2 s^{\gamma_1+\gamma_2+2}
e_g'(-s) = s^{\gamma_2} \int_0^s {s_1}^{\gamma_1}
\mbox{\textup{drivers}}_g(s_1) ds_1 + s^{\gamma_2} K_1',
\end{equation}
 we see that if $\gamma_2$ satisfies
\begin{equation}\label{gamma_2}
\gamma_2 = \frac{ (2-2g)(1-2g) - (\nu-1)^2
  (\gamma_1+1)(\gamma_1+2)}{(\gamma_1+1)(\nu-1)^2},
\end{equation}
one may solve equation (\ref{V.008}) for $e_g(s)$:
\begin{equation} \label{eg}
e_g(-s) = \frac{1}{(\nu-1)^2} s^{-\gamma_1-\gamma_2-2} \int_0^s
s_1^{\gamma_2} \int_0^{s_1} s_2^{\gamma_1} \mbox{\textup{drivers}}_g(s_2)
ds_2 ds_1 + K_1 s^{-\gamma_1-1} + K_2 s^{-\gamma_1-\gamma_2-2},
\end{equation}
where $K_1$ and $K_2$ are constants of integration.

Switching the order of integration in (\ref{eg})
\begin{equation} \label{eg2}
e_g(-s) = \frac{1}{(\nu-1)^2} s^{-\gamma_1-\gamma_2-2} \int_0^s
\int_{s_2}^s s_1^{\gamma_2} s_2^{\gamma_1} \mbox{\textup{drivers}}_g(s_2)
ds_1 ds_2 + K_1 s^{-\gamma_1-1} + K_2 s^{-\gamma_1-\gamma_2-2},
\end{equation}
one can compute the $s_1$ integral in (\ref{eg2}):
\begin{align} \label{eg3}
e_g(-s) &= \frac{1}{(\nu-1)^2} \frac{1}{\gamma_2+1} \left[
s^{-\gamma_1-1} \int_0^s
  s_2^{\gamma_1} \mbox{drivers}_g ds_2 - s^{-\gamma_1-\gamma_2-2} \int_0^s
  s_2^{\gamma_1+\gamma_2+2} \mbox{\textup{drivers}}_g ds_2 \right]
\\ \nonumber
& \phantom{=} + K_1 s^{-\gamma_1 -1} + K_2 s^{-\gamma_1 - \gamma_2 -2}.
\end{align}
Applying an integration by parts to the remaining integrals
(integrating the power of $s_2$ and differentiating the drivers
(with respect to $s_2$)):
\begin{align} \nonumber
e_g(-s) =& \frac{1}{(\nu-1)^2} \frac{1}{(\gamma_2+1)} \left[
  \left(\frac{1}{\gamma_1+1} - \frac{1}{\gamma_1+\gamma_2+2} \right)
  \mbox{\textup{drivers}}_g  +
  \frac{s^{-\gamma_1-1}}{\gamma_1+1} \int_0^s s_2^{\gamma_1+1}
  (\mbox{\textup{drivers}}_g)^{\bullet} z_0' ds_2 \right.
\\ \label{eg4}
& \left. - \frac{s^{-\gamma_1-\gamma_2-2}}{\gamma_1+\gamma_2+2} \int_0^s
  s_2^{\gamma_1+\gamma_2+2} (\mbox{\textup{drivers}}_g)^\bullet z_0'
  ds_2 \right]+ K_1 s^{-\gamma_1 -1} + K_2 s^{-\gamma_1 - \gamma_2 -2}.
\end{align}

We may now change variables and integrate with respect to $y =
z_0(s_2)$ in (\ref{eg4}):
\begin{align} \nonumber
e_g(-s) &= \frac{1}{(\nu-1)^2 (\gamma_2+1)} \left[
  \frac{\gamma_2+1}{(\gamma_1+1)(\gamma_1+\gamma_2+2)}
\mbox{\textup{drivers}}_g \right. \\
& \left. - \frac{1}{(\gamma_1+1)} \left( \frac{c_\nu z_0^\nu}{z_0-1}
\right)^{\gamma_1+1}
  \int^{z_0}_1 \left( \frac{y-1}{c_\nu y^\nu} \right)^{\gamma_1+1}
  (\mbox{\textup{drivers}}_g)^\bullet dy
\right. \\ & \hphantom{=} \nonumber \left. +
\frac{1}{\gamma_1+\gamma_2+2} \left( \frac{c_\nu z_0^\nu}{z_0-1}
  \right)^{\gamma_1+\gamma_2+2} \int^{z_0}_1 \left( \frac{y-1}{c_\nu y^\nu}
  \right)^{\gamma_1+\gamma_2+2} (\mbox{\textup{drivers}}_g)^\bullet dy \right]
\\ & \label{egfull}
\hphantom{=} + K_1 s^{-\gamma_1-1} + K_2 s^{-\gamma_1-\gamma_2-2},
\end{align}
where $K_1$ and $K_2$ are constants determined in either of the
following ways:  (1) by the requirement that $e_g$ be a locally
analytic function of $s$ or (2) by the evaluation of $e_g$ for low
values of $\nu$ through its combinatorial characterization.
\smallskip

\noindent Direct calculation shows that either choice of
$\gamma_1$ in (\ref{gamma_1}) produces formula (\ref{egform}) in
Theorem \ref{II.004thm}.

\subsection{Example $g=0$}

When $g=0$ we have
\begin{equation*}
\gamma_1 = {\gamma_1}^{(-)} = - \frac{\nu-2}{\nu-1}
\end{equation*}
then
\begin{equation*}
\gamma_2 = - \frac{\nu-2}{\nu-1}.
\end{equation*}
The driver when $g=0$ is just $\log(z_0)$. The two integrals in
(\ref{egfull}) may be evaluated separately:
\begin{align} \nonumber
\int^{z_0} (y-1)^{1/(\nu-1)} y^{-(2\nu-1)/(\nu-1)} dy &=
\int_1^{z_0} (1-y^{-1})^{1/(\nu-1)} y^{-2} dy
\\ \label{term1} &= \frac{\nu-1}{\nu} (1-z_0^{-1})^{\nu/(\nu-1)},
\end{align}
and
\begin{align} \nonumber
\int^{z_0} (y-1)^{2/(\nu-1)} y^{-(3\nu-1)/(\nu-1)} dy =&
\int_1^{z_0} ( 1-y^{-1})^{2/(\nu-1)} y^{-3} dy
\\ \nonumber=& \frac{(\nu-1)}{(\nu+1)} (1-z_0^{-1})^{(\nu+1)/(\nu-1)} z_0^{-1}
\\ \nonumber &
+ \frac{(\nu-1)}{(\nu+1)} \int_1^{z_0} (1-y^{-1})^{(\nu+1)/(\nu-1)} y^{-2}
dy
\\ \label{term2} =& \frac{(\nu-1)}{(\nu+1)} \left( \frac{z_0-1}{z_0^\nu}
\right)^{(\nu+1)/(\nu-1)} z_0^\nu
 + \frac{(\nu-1)^2}{2\nu(\nu+1)} \left( \frac{z_0-1}{z_0^\nu}
\right)^{2\nu /(\nu-1)} z_0^{2\nu} .
\end{align}
Plugging  (\ref{term1}) and (\ref{term2}) into (\ref{egfull}) with
$g=0$ yields:
\begin{align*}
e_0(-s) =& \frac{1}{2} \log(z_0) - \frac{\nu-1}{\nu} (z_0-1) +
\frac{1}{2} \frac{\nu-1}{\nu+1} (z_0-1) + \frac{1}{4}
\frac{(\nu-1)^2}{4\nu(\nu+1)} (z_0-1)^2
\\ =& \frac{1}{2} \log(z_0) + \frac{(\nu-1)^2}{4\nu(\nu+1)} (z_0 - 1)
\left(z_0 - \frac{3(\nu+1)}{\nu-1} \right) ,
\end{align*}
where $K_1$ and $K_2$ are both 0 by the analyticity condition.
This formula for $e_0$ agrees with the expression for $e_0(s)$
found using the equilibrium measure.

\subsection{Example:  $g=1$ \label{g=1section}}

The drivers for $e_1$ are
\begin{equation*}
\frac{z_1}{z_0} - \frac{1}{12} \frac{\partial^4}{\partial w^4}
\left( w^2 e_0(-w^{\nu-1} s ) \right),
\end{equation*}
evaluated at $w=1$.  As with the Forcing terms for the $z_g$ this
formula can be expressed as a rational function of $z_0$ with
poles at $z_0 = 0$ and $z_0 = \nu/(\nu-1)$.

We find that $g=1$ is the exceptional case for the method used in
formulas (\ref{V.006B})-(\ref{egfull}). We compute it directly
from equation (\ref{V.006}) with $g=1$. The integrating factors
are $(\gamma_1, \gamma_2) = (-1, -\nu/(\nu-1))$, differing from
the general choice of integrating factors (\ref{gamma_1}) and
(\ref{gamma_2}).

We find the integral formula:
\begin{align*}
e_1(-s) &= \frac{1}{(\nu-1)^2} s^{1/(\nu-1)} \int_0^s \int_0^s
s_1^{-\nu/(\nu-1)} s_2^{-1} \mbox{\textup{drivers}}_1(s_2) ds_2 ds_1
+ K_1 + K_2 s^{1/(\nu-1)}
\\
&= \frac{1}{(\nu-1)^2} s^{1/(\nu-1)} \int_0^s \int_{s_2}^s
s_1^{-\nu/(\nu-1)} s_2^{-1} \mbox{\textup{drivers}}_1(s_2) ds_1 ds_2
+ K_1 + K_2 s^{1/(\nu-1)}
\\
&= \frac{1}{(\nu-1)} \left[ s^{1/(\nu-1)} \int_0^s s_2^{-\nu/(\nu-1)}
  \mbox{\textup{drivers}}_1 ds_2 - \int_0^s s_2^{-1}
  \mbox{\textup{drivers}}_1 ds_2 \right] + K_1 + K_2 s^{1/(\nu-1)}
\\
&= \frac{1}{(\nu-1)} \left[ \left( \frac{z_0(s)-1}{c_\nu
    z_0(s)^\nu}\right)^{1/(\nu-1)} \int_1^{z_0(s)} \left( \frac{c_\nu
    y^\nu}{y-1} \right)^{\nu/(\nu-1)} \frac{(\nu - (\nu-1) y)}{c_\nu
    y^{\nu+1} } \mbox{\textup{drivers}}_1 dy
\right. \\  & \phantom{= \frac{1}{(\nu-1)}} \left.
- \int_1^{z_0(s)} \frac{(\nu-(\nu-1)y)}{y (y-1)}
\mbox{\textup{drivers}}_1 dy \right] + K_1 + K_2 s^{1/(\nu-1)}
\\
&= -\frac{1}{12} \log\left( \nu + (1-\nu) z_0 \right) + K_1 + K_2
s^{1/(\nu-1)};
\end{align*}
where we have switched the order of integration, computed the $s_1$
integral, and changed coordinates to integrals with respect to $y =
z_0(s_2)$.  The final step is a direct computation and is left to the
reader.

When $t=0$, $\log\left(Z_N(t)/Z_N(0)
\right) = 0$, therefore $e_1(0) = 0$ and so $K_1 = 0$.
We notice that if $\nu \neq 2$ then $K_2=0$ by the analyticity
condition, otherwise we will have to determine $K_2$. The
computation we will carry out in Section \ref{constants} shows
that when $\nu=2$,
\begin{equation*}
e_1(-s) = s + \mathcal{O}(s^2).
\end{equation*}
Our expression for $e_1(-s)$ has the
expansion (when $\nu=2$)
\begin{equation*}
-\frac{1}{12} \log\left[ 2 - z_0(s)\right] + K_2 s = (1 + K_2) s +
 \mathcal{O}(s^2),
\end{equation*}
therefore $K_2 = 0$.
This proves
formula (\ref{e1form}).

\subsection{Example:  $g=2$}

The drivers for $e_2(s)$ are
\begin{equation*}
\frac{z_2}{z_0} - \frac{1}{2} \frac{z_1^2}{z_0^2} - \frac{1}{12}
\frac{\partial^4}{\partial w^4} \left( e_1(-w^{\nu-1} s) \right) -
\frac{1}{360} \frac{\partial^6}{\partial w^6} \left( w^2
e_0(-w^{\nu-1} s) \right),
\end{equation*}
evaluated at $w=1$.  As with the Forcing terms for the $z_g$ and
the drivers for $e_1$ this expression can be expressed as a
rational function of $z_0$ with poles at $z_0 = 0$ and $z_0 =
\nu/(\nu-1)$.

For $g=2$ we find that
\begin{equation*}
\gamma_1 = \gamma_1^{(+)} = - \frac{\nu+1}{\nu-1}
\end{equation*}
and
\begin{equation*}
\gamma_2 = -\frac{\nu}{\nu-1} .
\end{equation*}

We compute the integrals in equation (\ref{egfull}),
\begin{align*}
e_2(-s) =& \frac{1}{2880} (\nu-1) (z_0-1) \left( \nu - (\nu-1) z_0
\right)^{-5} \left( (-\nu^3 + 5 \nu^4 + 8 \nu^5 ) \right. \\ &
\left. + (-\nu^2 + 41 \nu^3 - 24 \nu^4 - 16 \nu^5 ) z_0 + (44\nu -
89\nu^2 + 54\nu^3 - 17\nu^4 + 8 \nu^5 )z_0^2 \right. \\ & \left. +
(-12 -12\nu +108\nu^2 - 132 \nu^3 + 48 \nu^4 ) z_0^3 + (-12 + 48
\nu - 72 \nu^2 + 48 \nu^3 - 12 \nu^4) z_0^4 \right)
\\ & + K_1 s^{2/(\nu-1)} + K_2 s^{3/(\nu-1)}.
\end{align*}
We notice that $K_1$ and $K_2$ will be zero by the analyticity
condition unless $\nu = 2, 3$, or $4$ in which case they will have
to be determined by some other means such  as that
illustrated in the next subsection.

\subsection{Evaluating the constants of integration \label{constants} }

We will now outline a method for determining the constants of
integration in formulas (\ref{egform}) and (\ref{e1form})
of Theorem \ref{II.004thm}.

We know, from \cite{EM}, that $e_g(-s)$ is analytic in $s$ in a
neighborhood of $s=0$, therefore we notice that for many values of
$\gamma_1$ both $K_1$ and $K_2$ will vanish to preserve
analyticity. However there are some values for which they will
not: Values of $\nu$ and $g$ which conspire to make
$(2g-2)/(\nu-1)$ or $ (2g-1)/(\nu-1)$ be positive integers. To
find these constants, when necessary, one may rely on the
combinatorial interpretation of the Taylor coefficients of $e_g$
 described in Theorem \ref{thm:01}.  Here we outline the method
that we have used to evaluate these constants for $g\leq 3$.

The set of maps $D$ having vertex set $K_0(D)$ of fixed
cardinality at specified vertices can be placed in 1-1
correspondence with a class of subgroups of the permutation group
$\mathcal{S}_d$ where $d$, twice the number of edges, is
determined by the cardinality and valences of $K_0(D)$ \cite{BI2}.
For simplicity and also because it is the case of relevance for
us, we take $K_0(D)$ to consist of $n$ vertices each of valence
$2\nu$. For this class $d=2\nu n$ which is the cardinality of the
set of darts of $K(D)$. A \emph{dart} is an oriented edge.
Equivalently, we may define the abstract set of darts associated
to this class of maps as
\begin{equation*}
\Omega = \Omega_1 \cup \Omega_2,
\end{equation*}
where
\begin{equation*}
\Omega_1 = \left\{ (v, e):
\mbox{ $v$ is a vertex and $e$ is an edge with two distinct vertices
  one of them being $v$}
\right\},
\end{equation*}
and
\begin{equation*}
\Omega_2 = \left\{ (v, e, \pm):
\mbox{ $v$ is a vertex and $e$ is an edge with a single vertex $v$}
 \right\}.
\end{equation*}
The element $(v, e) \in \Omega_1$ represents the dart based at $v$ and
going along $e$.  The element $(v, e, \pm) \in \Omega_2$ represents
the dart based at $v$ going along $e$ in the counterclockwise
(resp. clockwise) orientation.
For each edge there are two darts, therefore $\big| \Omega \big| = d$
and
we can think of $\mathcal{S}_d$ as acting by permutations on the
set $\Omega$.

Given a map, $(K(D),[\iota])$ with $K_0(D)$ specified as above, we
define a subgroup of $\mathcal{S}_d$ generated by two permutations
$\langle \sigma, \tau \rangle$.
The orientation on $X$ induces (via $[\iota]$) a cyclic ordering on
the darts attatched to each vertex; the first permutation $\sigma$ is
given by this action.  Explicitly, $\sigma$ maps the element $(v, e)
\in \Omega_1$ to the element $(v, \tilde{e}) \in \Omega_1$ or $(v,
\tilde{e}, \pm) \in \Omega_2$
 where $\tilde{e}$ is the
edge counter clockwise in the orientation at $v$ from $e$.
Likewise $\sigma$ maps the element $(v, e, \pm) \in \Omega_2$ to the
element $(v, \tilde{e}) \in \Omega_1$ or $(v, \tilde{e}, \pm) \in
\Omega_2$ where $\tilde{e}$ is the edge counter clockwise in the
orientation at $v$ from $e$.
  The second permutation $\tau$, is given explicitly as
the permutation which acts on $\Omega_1$ by sending $(v, e)$ to
$(\tilde{v}, e)$ where  $\tilde{v}$ is the other endpoint of $e$;
$\tau$ acts on $\Omega_2$ by sending $(v, e, \pm)$ to $(v, e, \mp)$.

>From these descriptions one sees that: $\sigma$ is a product of
disjoint $2\nu$-cycles, with each cycle corresponding to a unique
vertex in $K_0(D)$; and that $\tau$ is a product of disjoint
2-cycles, with each 2-cycle corresponding to a unique edge in
$K_1(D)$.

Conversely, given a subgroup presented as above which also has the
property that the group acts transitively on $\Omega$, one may
construct a unique map. The transitivity condition insures that
the underlying map is connected. The permutation $\tau$ determines
how the vertices are connected through edges to define the graph.
The permutation $\sigma$ gives the orientation of the edges about
each vertex. Together these two permutations determine the surface
$X$ and the embedding class $\left[ \iota \right]$.

The punchline is that what we now have is an algorithm for computing
the coefficient of $s^n / n!$ in  $e_g(-s)$ for finite $g$, $\nu$, and $n$:
Let $d = 2 \nu n$.  Fix $\sigma$ to be a permutation formed by a
disjoint product of $n$
$2\nu$-cycles in $\mathcal{S}_d$.  Then we choose each disjoint
 product of $\nu n$ 2-cycles, $\tau$ in $\mathcal{S}_d$.  Check if
$(\sigma, \tau)$ is connected (by verifying that the orbit of $\langle
\sigma, \tau \rangle \cdot 1$ is all $d$ letters). If $(\sigma, \tau)$
is connected compute the genus by Euler's formula
$(1-\nu) n + F = \chi = 2-2g$, where
$F$ (the number of faces) is given by the number of cycles in $\sigma
\circ \tau$.  The details of this calculation together with some
examples and a generalization to the case of unoriented maps are
in a forthcoming paper \cite{Pierce_3}.

The algorithm gives the following values for the coefficient of
$s^j/j!$ in $e_g(-s)$:
\begin{center}
\begin{tabular}{|lccr|}\hline
$g=1$ &         & $j=0$ & 0 \\
$g=1$ & $\nu=2$ & $j=1$ & 1  \\
\hline
$g=2$ & $\nu=2$ & $j=2$ & 0 \\
$g=2$ & $\nu=2$ & $j=3$ & 1440 \\
$g=2$ & $\nu=3$ & $j=1$ & 0 \\
$g=2$ & $\nu=4$ & $j=1$ & 21 \\
\hline
$g=3$ & $\nu=2$ & $j=4$ & 0 \\
$g=3$ & $\nu=2$ & $j=5$ & 58060800 \\
$g=3$ & $\nu=3$ & $j=2$ & 0 \\
$g=3$ & $\nu=5$ & $j=1$ & 0 \\
$g=3$ & $\nu=6$ & $j=1$ & 1485 \\
\hline
\end{tabular}
\end{center}
Then $K_1$ and $K_2$ are chosen so that the $j$'th coefficient of
$e_g(-s)$ matches these numbers.

\subsection{Multiple times \label{multtimes}}

The constructions carried out in this paper extend to multiple
even time parameters; this is a reflection of the commutativity of
the underlying flows in the Toda Lattice hierarchy. However, the
expressions found are more complicated and are less easily reduced
to closed form than the monic time cases we have considered thus
far.

 For brevity we
will show how our results extend to two times: $t_{2\nu_1}$ and
$t_{2\nu_2}$, where $\nu_1$ and $\nu_2$ are positive integers.
Theorem \ref{II.002} becomes
\begin{thm}\label{V.003thm}
In the limit as $k \to \infty$, $b_k(\xi_{2\nu_1},
\xi_{2\nu_2})^2$ has a valid asymptotic expansion of the form
\begin{equation*}
b_k(\xi_{2\nu_1}, \xi_{2\nu_2})^2 \simeq k ( z_0(s_{2\nu_1},
s_{2\nu_2}) + \frac{1}{k^2} z_1(s_{2\nu_1}, s_{2\nu_2}) +
\frac{1}{k^4} z_2(s_{2\nu_1}, s_{2\nu_2}) + \dots)
\end{equation*}
where $s_{2\nu_i} = 2k^{\nu_i-1} \xi_{2\nu_i}$.  The terms of this
expansion are determined by the following partial differential
scheme:
\begin{equation*}
f_{s_{2\nu_i}} = c_{\nu_i} f^{\nu_i} f_w + \frac{1}{k^2}
F_1^{(\nu_i)}(f, f_w, f_{ww}, f_{www} ) + \dots + \frac{1}{k^{2g}}
F_g^{(\nu_i)}(f, f_w, f_{w^{(2)}}, \dots, f_{w^{(2g+1)}} )+ \dots
\bigg|_{\mbox{evaluated at $w=1$}};
\end{equation*}
where
\begin{align*}
f(s_{2\nu_1}, s_{2\nu_2}; w) &= f_0(s_{2\nu_1}, s_{2\nu_2}; w) +
\frac{1}{k^2} f_1(s_{2\nu_1}, s_{2\nu_2}; w) + \frac{1}{k^4}
f_2(s_{2\nu_1}, s_{2\nu_2}; w) + \dots, \mbox{ and }f_g \mbox{ has the form } \\
f_g(s_{2\nu_1}, s_{2\nu_2}; w) &= w^{1-2g} z_g( w^{\nu_1-1}
s_{2\nu_1}, w^{\nu_2-1} s_{2\nu_2} ).
\end{align*}
 This partial differential scheme yields a hierarchy of partial
 differential equations in the same manner as in Theorem
 \ref{II.002}.  The functionals $F_g^{(\nu_i)}$ are identical to the
 ones found in Theorem \ref{II.002}.

\end{thm}

The $z_g(s_{2\nu_1}, s_{2\nu_2})$ play the role of auxiliary
functions in computing $e_g(-s_{2\nu_1}, -s_{2\nu_2})$.
The differential equation determining $e_g(-s_{2\nu_1}, -s_{2\nu_2})$
remains largely unchanged from (\ref{eg_equation}) because the
construction in section \ref{diff_eq_eg} applies for the multi-time
case as well:
\begin{thm}
The $g$'th equation in the hierarchy of equations governing
$e_g(t_{2\nu_1}, t_{2\nu_2})$ is
\begin{align} \label{two_time_eg}
&\frac{\partial^2}{\partial w^2} \left[ w^{2-2g} e_g(-w^{\nu_1-1}
  s_{2\nu_1}, - w^{\nu_2-1}s_{2\nu_2} ) \right]_{w=1}
\\ \nonumber
&=
 - \sum_{n=1}^g \frac{2}{(2n+2)!} \frac{\partial^{2n+2} }{\partial
  w^{2n+2}} \left[ w^{2-2(g-n)} e_{g-n}(-w^{\nu_1-1} s_{2\nu_1},
  -w^{\nu_2-1} s_{2\nu_2}) \right]   \\ \nonumber
& \phantom{=}
+\mbox{the $k^{-2g}$ term of} \quad \log\left[ \sum_{n=0}^\infty
  \frac{1}{k^{2n}} z_n(s_{2\nu_1}, s_{2\nu_2}) \right].
\end{align}
 \end{thm}

Equation (\ref{two_time_eg}) determines $e_g(-s_{2\nu_1},
-s_{2\nu_2})$ from a second order partial differential equation of
$e_g$ with forcing terms depending on $e_n, \, n<g, \; z_n, \, n\leq
g,$ and their derivatives with respect to $s_{2\nu_1}$ and
$s_{2\nu_2}$.

\begin{thm} \label{two_time_eg_thm}
The RHS of (\ref{two_time_eg}) will henceforth be denoted by
$\mbox{\textup{drivers}}_g(s_{2\nu_1}, s_{2\nu_2})$.
The solution of (\ref{two_time_eg}) may be represented as
\begin{align} \label{two_time_eg_6}
e_g(-s_{2\nu_1}, -s_{2\nu_2}) &= (\nu_1-1)^{-1}
\\ \nonumber & \phantom{=} \nonumber
\left[ s_{2\nu_1}^{-(1-2g)/(\nu_1-1)}
\int_0^{s_{2\nu_1}} \left(
  \hat{s}_{2\nu_1} \right)^{(2-\nu_1-2g)/(\nu_1-1)}
  \mbox{\textup{drivers}}_g(\hat{s}_{2\nu_1}, u
       {\hat{s}}_{2\nu_1}^{(\nu_2-1)/(\nu_1-1)}) d\hat{s}_{2\nu_2}
\right. \\ \nonumber & \phantom{=} \nonumber \left.
- s_{2\nu_1}^{-(2-2g)/(\nu_1-1)} \int_0^{s_{2\nu_1}} \left(
\hat{s}_{2\nu_1} \right)^{(3 - \nu_1 - 2g)/(\nu_1-1)}
\mbox{\textup{drivers}}_g(\hat{s}_{2\nu_1}, u
     {\hat{s}}_{2\nu_1}^{(\nu_2-1)/(\nu_1-1)}) d\hat{s}_{2\nu_1} \right]
\\ \nonumber & \phantom{=} \nonumber
+ K_1(u) s_{2\nu_1}^{(2-2g)/(\nu_1-1)}  + K_2(u)
s_{2\nu_1}^{(1-2g)/(\nu_1-1)},
\end{align}
where $u=s_{2\nu_2} s_{2\nu_1}^{-(\nu_2-1)/(\nu_1-1)}$, and where $K_1(u)$
and $K_2(u)$ are analytic functions of $u$ in a neighborhood of $u=0$.
\end{thm}

To prove Theorem \ref{two_time_eg_thm}:  first we
will show that $z_0(s_{2\nu_1}, s_{2\nu_2})$ is given implicitly
as the solution of an algebraic equation with coefficients depending
on $s_{2\nu_1}$ and $s_{2\nu_2}$.  Then we demonstrate that
$z_g(s_{2\nu_1}, s_{2\nu_2})$ are functions of $z_0(s_{2\nu_1},
s_{2\nu_2})$ and $z_0(s_{2\nu_1}, 0)$.  Finally we compute
$e_g(-s_{2\nu_1}, -s_{2\nu_2})$ by integrating
 the partial differential equation
in Theorem \ref{two_time_eg}.

The first order terms in the hierarchy of Theorem \ref{V.003thm}
are the pair of equations
\begin{align*}
\frac{dz_0}{ds_{2\nu_1}} &= c_{\nu_1} z_0^{\nu_1} \left( z_0 +
(\nu_1-1) s_{2\nu_1} \frac{dz_0}{ds_{2\nu_1}} + (\nu_2 - 1) s_{2\nu_2}
\frac{dz_0}{ds_{2\nu_2}} \right) \\
\frac{dz_0}{ds_{2\nu_2}} &= c_{\nu_2} z_0^{\nu_2} \left( z_0 + (\nu_1
-1) s_{2\nu_1} \frac{dz_0}{ds_{2\nu_1}} + (\nu_2-1) s_{2\nu_2}
\frac{dz_0}{ds_{2\nu_2}} \right);
\end{align*}
or in vector notation
\begin{equation} \label{two_time_0}
\mathbf{M}
\begin{pmatrix} \frac{dz_0}{ds_{2\nu_1}} \\ \frac{dz_0}{ds_{2\nu_2}} \end{pmatrix}
= \begin{pmatrix} c_{\nu_1} z_0^{\nu_1+1} \\ c_{\nu_2}
  z_0^{\nu_2+1} \end{pmatrix},
\end{equation}
where
\begin{equation*}
\mathbf{M} = \begin{pmatrix} 1 - c_{\nu_1} (\nu_1-1) s_{2\nu_1}
  z_0^{\nu_1}  & - c_{\nu_1} (\nu_2-1) s_{2\nu_2} z_0^{\nu_1} \\
- c_{\nu_2} (\nu_1-1) s_{2\nu_1} z_0^{\nu_2} & 1 - c_{\nu_2} (\nu_2-1)
s_{2\nu_2} z_0^{\nu_2} \end{pmatrix}.
\end{equation*}
Invert $\mathbf{M}$ in (\ref{two_time_0}) to find a pair of ordinary
differential equations for $z_0$:
\begin{equation*}
\frac{dz_0}{ds_{2\nu_i} } = - \frac{c_{\nu_i} z_0^{\nu_i +1} }{ -1 +
  c_{\nu_1} (\nu_1 -1) s_{2\nu_1} z_0^{\nu_1} + c_{\nu_2}
  (\nu_2-1)s_{2\nu_2} z_0^{\nu_2} },
\end{equation*}
which, with initial condition $z_0(0, 0)=1$,  is satisfied implicitly by
the solution of
\begin{equation} \label{two_time_constraint}
1 = z_0 - c_{\nu_1} s_{2\nu_1} z_0^{\nu_1} - c_{\nu_2} s_{2\nu_2} z_0^{\nu_2}
\end{equation}
which is regular at $(s_{2\nu_1}, s_{2\nu_2}) = (0, 0)$.

The $g^{\mbox{th}}$ pair of equations in the hierarchy of partial
differential equations  in Theorem \ref{V.003thm} is
\begin{equation} \label{two_time_2}
\mathbf{M} \begin{pmatrix} \frac{dz_g}{ds_{2\nu_1}}
  \\ \frac{dz_g}{ds_{2\nu_2}} \end{pmatrix} = z_g \mathbf{G}_g +
\mathbf{F}_g;
\end{equation}
where
\begin{align*}
\mathbf{G}_g &= \begin{pmatrix} c_{\nu_1} (\nu_1+1-2g) z_0^{\nu_1} +
  c_{\nu_1} \nu_1 (\nu_1 -1) s_{2\nu_1} z_0^{\nu_1 - 1}
  \frac{dz_0}{ds_{2\nu_1}} + c_{\nu_1} \nu_1 (\nu_2-1) s_{2\nu_2}
  z_0^{\nu_2-1} \frac{dz_0}{ds_{2\nu_2}} \\
c_{\nu_2} (\nu_2+1-2g) z_0^{\nu_2} + c_{\nu_2} \nu_2 (\nu_1-1)
s_{2\nu_1} z_0^{\nu_2-1} \frac{dz_0}{ds_{2\nu_1}} + c_{\nu_2} \nu_2
(\nu_2-1) s_{2\nu_2} z_0^{\nu_2-1} \frac{dz_0}{ds_{2\nu_2} }
\end{pmatrix}
\\
\mathbf{F}_g &= \begin{pmatrix} \mbox{\textup{Forcing}}_g^{(\nu_1)}
  \\ \mbox{\textup{Forcing}}_g^{(\nu_2)} \end{pmatrix} \bigg|_{w=1},
\end{align*}
and where $\mbox{\textup{Forcing}}_g^{(\nu_i)}$ is given by
(\ref{Forcing}) with
\begin{equation*}
f_g(s_{2\nu_1}, s_{2\nu_2}; w) = w^{1-2g} z_g( w^{\nu_1 -1}
s_{2\nu_1}, w^{\nu_2-1} s_{2\nu_2}) .
\end{equation*}

It is useful at this stage to make the following change of
variables: let $z_0(s_{2\nu_1}, 0) = y_0$ and denote
$z_0(s_{2\nu_1}, s_{2\nu_2})$ as $z_0$. The constraint equation
(\ref{two_time_constraint}) becomes:
\begin{align}
\label{two_time_y0} 1 &= y_0 - c_{\nu_1} s_{2\nu_1} y_0^{\nu_1} \\
\label{two_time_z0} 1 &= z_0 - c_{\nu_1} s_{2\nu_1} z_0^{\nu_1} -
c_{\nu_2} s_{2\nu_2} z_0^{\nu_2},
\end{align}
and one can solve equations (\ref{two_time_y0}) and
(\ref{two_time_z0}) for $(s_{2\nu_1}, s_{2\nu_2})$ as functions of
$(y_0, z_0)$,
\begin{align} \label{two_time_s1}
s_{2\nu_1} &= ( y_0 - 1 ) ( c_{\nu_1} y_0^{\nu_1} )^{-1}
\\ \label{two_time_s2}
s_{2\nu_2} &= ( y_0^{\nu_1} (z_0 -1) - z_0^{\nu_1} ( y_0 - 1) ) (
c_{\nu_2} y_0^{\nu_1} z_0^{\nu_2} )^{-1},
\end{align}
and then differentiate equation (\ref{two_time_y0} and
\ref{two_time_z0}) with respect to $s_{2\nu_1}$ and $s_{2\nu_2}$
and solve for $\frac{dy_0}{ds_{2\nu_1}}$,
$\frac{dz_0}{ds_{2\nu_1}}$ and $\frac{dz_0}{ds_{2\nu_2}}$ as
functions of $(y_0, z_0)$:
\begin{align*}
\frac{dy_0}{ds_{2\nu_1}} &= c_{\nu_1} y_0^{\nu_1+1} ( \nu_1 -
(\nu_1-1) y_0 )^{-1} \\
\frac{dy_0}{ds_{2\nu_2}} &= 0 \\
\frac{dz_0}{ds_{2\nu_1}} &= c_{\nu_1} z_0^{\nu_1} ( 1 - \nu_1
c_{\nu_1} s_{2\nu_1} z_0^{\nu_1-1} - \nu_2 c_{\nu_2} s_{2\nu_2}
z_0^{\nu_2-1} )^{-1} \\
\frac{dz_0}{ds_{2\nu_2}} &= c_{\nu_2} z_0^{\nu_2} ( 1 - \nu_1
c_{\nu_1} s_{2\nu_1} z_0^{\nu_1-1} - \nu_2 c_{\nu_2} s_{2\nu_2}
z_0^{\nu_2-1} )^{-1}.
\end{align*}
Changing variables in the system of differential equations
(\ref{two_time_2}) to differential equations for $z_g$ as a
function of $(y_0, z_0)$; where $y_0$ evolves from $1$ to
$z_0(s_{2\nu_1}, 0)$ and $z_0$ from $z_0(s_{2\nu_1}, 0)$ to
$z_0(s_{2\nu_1}, s_{2\nu_2} ) $, equation (\ref{two_time_2})
becomes
\begin{equation}
\mathbf{M} \mathbf{C}
\begin{pmatrix} \frac{dz_g}{dy_0} \\ \frac{dz_g}{dz_0} \end{pmatrix} =
z_g \mathbf{G}_g + \mathbf{F}_g,
\end{equation}
where
\begin{equation*}
\mathbf{C} = \begin{pmatrix} c_{\nu_1} y_0^{\nu_1+1} (\nu_1 -
  (\nu_1-1) y_0 )^{-1} &
c_{\nu_1} z_0^{\nu_1} ( 1- \nu_1 c_{\nu_1} s_{2\nu_1} z_0^{\nu_1-1} -
\nu_2 c_{\nu_2} s_{2\nu_2} z_0^{\nu_2-1} )^{-1} \\
0 & c_{\nu_2} z_0^{\nu_2} ( 1- \nu_1 c_{\nu_1} s_{2\nu_1}
z_0^{\nu_1-1} - \nu_2 c_{\nu_2} s_{2\nu_2} z_0^{\nu_2-1} )^{-1}
\end{pmatrix},
\end{equation*}
and $s_{2\nu_1}$ and $s_{2\nu_2}$ are given by equations
(\ref{two_time_s1}) and (\ref{two_time_s2}).

If $\mathbf{M} \mathbf{C}$ is invertible then $z_g$ can be found by
integrating the differential equation:
\begin{equation} \label{two_time_zg}
\frac{dz_g}{dz_0}= z_g \left[ \mathbf{C}^{-1} \mathbf{M}^{-1}
  \mathbf{G}_g \right]_2 + \left[ \mathbf{C}^{-1} \mathbf{M}^{-1}
  \mathbf{F}_g \right]_2,
\end{equation}
where $[ \mathbf{V} ]_2$ denotes the second component of the vector
$\mathbf{V}$.
The initial condition is that $z_g(y_0, z_0=y_0)$ agrees with
Theorem \ref{II.003thm} with $\nu = \nu_1$.


We will now prove Theorem \ref{two_time_eg_thm}.  Start by
expanding the LHS of (\ref{two_time_eg}):
\begin{align} \label{two_time_eg_2}
(2-2g)(1-2g)e_g - (\nu_1 -1)(\nu_1+2-4g)s_{2\nu_1} e_{g s_{2\nu_1}} - (\nu_2
  -1)(\nu_2 +2 - 4g) s_{2\nu_2} e_{g s_{2\nu_2}}
& \\ \nonumber
+ (\nu_1-1)^2
  s_{2\nu_1}^2 e_{g s_{2\nu_1} s_{2\nu_1}}
+ 2 (\nu_1 -1)(\nu_2-1)
  s_{2\nu_1} s_{2\nu_2} e_{g s_{2\nu_1} s_{2\nu_2}} + (\nu_2-1)^2
  s_{2\nu_2}^2 e_{g s_{2\nu_2} s_{2\nu_2}} &=
  \mbox{\textup{drivers}}_g(s_{2\nu_1}, s_{2\nu_2}).
\end{align}
Changing variables to
\begin{equation*}
(\hat{s}_{2\nu_1}, u) = ( s_{2\nu_1}, s_{2\nu_2}
  s_{2\nu_1}^{-(\nu_2-1)/(\nu_1-1)} )
\end{equation*}
induces
\begin{equation*}
\begin{pmatrix}
\hat{s}_{2\nu_1} \frac{\partial}{\partial \hat{s}_{2\nu_1}} \\
u \frac{\partial}{\partial u}
\end{pmatrix}
=
\begin{pmatrix}
s_{2\nu_1} \frac{\partial}{\partial s_{2\nu_1}} + \frac{(\nu_2-1)}{(\nu_1-1)} s_{2\nu_2}
\frac{\partial}{\partial s_{2\nu_2}} \\
s_{2\nu_2} \frac{\partial}{\partial s_{2\nu_2}}
\end{pmatrix}.
\end{equation*}
Equation (\ref{two_time_eg_2}) may be rewritten in these new
variables as:
\begin{equation} \label{two_time_eg_3}
(\nu_1-1)^2 \left( \hat{s}_{2\nu_1} \frac{\partial}{\partial
    \hat{s}_{2\nu_1}} + \frac{(1-2g)}{(\nu_1-1)} \right)
\left( \hat{s}_{2\nu_1} \frac{\partial}{\partial \hat{s}_{2\nu_1}} +
    \frac{(2-2g)}{(\nu_1-1)} \right) e_g =
    \mbox{\textup{drivers}}_g(\hat{s}_{2\nu_1}, u
    \hat{s}_{2\nu_1}^{(\nu_2-1)/(\nu_1-1)}),
\end{equation}
whose LHS is identical to the differential equation (\ref{V.006})
for $e_g(-s_{2\nu_1})$.

Equation (\ref{two_time_eg_3}) can be integrated to give
\begin{align} \label{two_time_eg_4}
e_g(-s_{2\nu_1}, -s_{2\nu_2}) &= (\nu_1-1)^{-2}
s_{2\nu_1}^{-(2-2g)/(\nu_1-1)}
\int_0^{s_{2\nu_1}} \left( s'_{2\nu_1}
\right)^{(2-\nu_1)/(\nu_1-1)} \int_0^{s'_{2\nu_1}} \left( s''_{2\nu_1}
\right)^{(2-\nu_1 - 2g)/(\nu_1-1)}
\\ & \phantom{=(\nu_1-1)^{-2}
s_{2\nu_1}^{-(2-2g)/(\nu_1-1)}  } \nonumber \cdot
\mbox{\textup{drivers}}_g(
s''_{2\nu_1}, u {s''}_{2\nu_1}^{(\nu_2-1)/(\nu_1-1)} ) ds''_{2\nu_1}
ds'_{2\nu_1}
\\ & \phantom{=} \nonumber
+ K_1(u) s_{2\nu_1}^{(2-2g)/(\nu_1-1)} + K_2(u)
s_{2\nu_1}^{(1-2g)/(\nu_1-1)},
\end{align}
where $u = s_{2\nu_2} s_{2\nu_1}^{-(\nu_2-1)/(\nu_1-1)}$ and $K_1(u)$
and $K_2(u)$ are functions of $u$ only.

Switch the order of integration in (\ref{two_time_eg_4}),
\begin{align} \label{two_time_eg_5}
e_g(-s_{2\nu_1}, -s_{2\nu_2}) &= (\nu_1-1)^{-2}
s_{2\nu_1}^{-(2-2g)/(\nu_1-1)}
\int_0^{s_{2\nu_1}}
\int_{s''_{2\nu_1}}^{s_{2\nu_1}}
\left( s'_{2\nu_1}
\right)^{(2-\nu_1)/(\nu_1-1)} \left( s''_{2\nu_1}
\right)^{(2-\nu_1 - 2g)/(\nu_1-1)}
\\ & \phantom{=(\nu_1-1)^{-2}
s_{2\nu_1}^{-(2-2g)/(\nu_1-1)}  } \nonumber \cdot
\mbox{\textup{drivers}}_g(
s''_{2\nu_1}, u {s''}_{2\nu_1}^{(\nu_2-1)/(\nu_1-1)} ) ds'_{2\nu_1}
ds''_{2\nu_1}
\\ & \phantom{=} \nonumber
+ K_1(u) s_{2\nu_1}^{(2-2g)/(\nu_1-1)} + K_2(u)
s_{2\nu_1}^{(1-2g)/(\nu_1-1)},
\end{align}
and carry out the $s'_{2\nu_1}$ integral in (\ref{two_time_eg_5})
to find
\begin{align*}
e_g(-s_{2\nu_1}, -s_{2\nu_2}) &= (\nu_1-1)^{-1}
\\ & \phantom{=} \nonumber
\left[ s_{2\nu_1}^{-(1-2g)/(\nu_1-1)}
\int_0^{s_{2\nu_1}} \left(
  s''_{2\nu_1} \right)^{(2-\nu_1-2g)/(\nu_1-1)}
  \mbox{\textup{drivers}}_g(s''_{2\nu_1}, u
       {s''}_{2\nu_1}^{(\nu_2-1)/(\nu_1-1)}) ds''_{2\nu_2}
\right. \\ & \phantom{=} \nonumber \left.
- s_{2\nu_1}^{-(2-2g)/(\nu_1-1)} \int_0^{s_{2\nu_1}} \left(
s''_{2\nu_1} \right)^{(3 - \nu_1 - 2g)/(\nu_1-1)}
\mbox{\textup{drivers}}_g(s''_{2\nu_1}, u
     {s''}_{2\nu_1}^{(\nu_2-1)/(\nu_1-1)}) ds''_{2\nu_1} \right]
\\ & \phantom{=} \nonumber
+ K_1(u) s_{2\nu_1}^{(2-2g)/(\nu_1-1)}  + K_2(u)
s_{2\nu_1}^{(1-2g)/(\nu_1-1)}.
\end{align*}
This proves formula (\ref{two_time_eg_6}) of Theorem
\ref{two_time_eg_thm}.

\section{Conclusions}

In this paper we have made a detailed study of the coefficients
$e_g(x,t)$ in the asymptotic expansion of the logarithm of the
random matrix partition function (\ref{I.002}) for a single
non-trivial time parameter, $t = t_{2 \nu}$ and $x=k/N$ near $1$.
These analytic coefficients are generating functions for the
enumeration of \emph{g-maps}. In particular,

\begin{enumerate}
\item We derived a hierarchy of differential equations for these
generating functions.

\item We described a procedure for solving these differential
equations.

\item Along the way to deriving the hierarchy we also derive an
hierarchy of forced Burgers type equations for the auxiliary
coefficients which we denote $z_g(s)$; these are of combinatorial
interest in their own right in that
$$
\frac{\partial^n}{\partial s^n} z_g(0)= ^\sharp \{\mbox{two-legged
$g$-maps with $n$ $2\nu$-valent vertices }\}.
$$
A \emph{leg} is an edge emerging from a univalent vertex; so that
the leg is the only edge incident to that vertex.

\item We have calculated explicit formulae for $e_g(-s)$ for small
values of $g$. The $s$-derivatives, of sufficiently large order,
turn out to be rational functions of the endpoints squared,
$\beta^2 = 4z_0$. The endpoints referred to here are the endpoints
of the support of the associated equilibrium measure. Moreover,
the coefficients appearing in these expressions are rational
constants.

\end{enumerate}

The random matrix partition functions and their relations to
graphical enumeration through diagrammatic expansion offer
powerful tools for bringing methods of complex analysis to bear on
fundamental questions in diverse fields such as statistical
mechanics \cite{Difrancesco95} and combinatorics \cite{Witten,
BIZ}. However, real progress in this regard has been hampered by
the fact that, up till now, these connections have been based on
formal procedures and conjectures.

The results in this paper place many of these connections, for the
large $N$ expansion of the UE partition function, on a rigorous
foundation. Moreover, the methods presented here, based on
continuum limits of the Toda Lattice hierarchy, have yielded novel
and effective procedures for explicitly calculating the relevant
asymptotic generating functions. We hope that these results will
help to spur renewed application of complex analytic methods in
problems of statistical physics and combinatorics. Some results in
this direction will appear in future work \cite{Pierce,Pierce_3}.

These analytical tools also point the way to novel combinatorial
reults. The calculations mentioned in the last item of the list
above, strongly motivate the
\bigskip

\textbf{Conjecture:} The $s$-derivatives of sufficiently high
order of the generating functions, $e_g$, for fixed genus with
arbitrary vertex valence number, can be expressed as rational
functions of the endpoints of support of the equilibrium measure.
Moreover, the coefficients appearing in these expressions are
rational constants.
\medskip

\noindent As far as we know, such a conjecture has not appeared in
the literature on combinatorics of maps.
\medskip

Another manuscript in preparation \cite{ErMcLNew} rigorously
establishes a nonlocal representation for the $e_g(1,t)$. In the
physics literature such representations are referred to as
\emph{loop equations} \cite{Ambjorn}. We expect this to provide
elegant methods that, together with the present manuscript, could
enable us to prove the previous conjecture. Moreover, the
generalizations of this loop equation together with the results of
section \ref{multtimes} can help to guide the characterization of
$e_g(\textbf{t})$ as a function of multiple times. The derivation
of explicit closed form expressions for $e_g(\textbf{t})$ will
have relevance to a number of other current research programs in
the statistical mechanics of combinatorial analysis such as random
graphs, random tilings and polynuclear growth models.
\bigskip
\bigskip

\noindent {\bf Acknowledgements:} V. U. Pierce would like to thank
The University of Arizona, Brandeis University, Midwestern State
University, and The Ohio State University for their hospitality
and support; and Mark Adler and Yuji Kodama for many helpful
discussions and their encouragement.


\appendix
\section{\label{A}}

In this appendix we show how the expansion (\ref{intexp}) for the
integral (\ref{4_25}) was derived.

Define
\begin{equation} \label{d_j}
d_j = \int_\beta^\lambda s^{2j} \sqrt{s^2 - \beta^2} ds,
\end{equation}
then (\ref{4_25}) is
\begin{equation*}
\int_{\beta}^\lambda h(s) \sqrt{s^2 - \beta^2} ds = \frac{1}{x}
\left( d_0 + \sum_{j=0}^{\nu-1} h_j d_j \right).
\end{equation*}

Integration by parts ( $u = s^{2j-1}$ and $dv = s \sqrt{s^2 -
\beta^2}$) of $d_j$ gives the equation
\begin{equation} \label{d_j_step1}
d_j = \frac{1}{3} \lambda^{2j-1} (\lambda^2 - \beta^2)
\sqrt{\lambda^2 - \beta^2} - \int \frac{(2j-1)}{3} s^{2j-2} (s^2 -
\beta^2) \sqrt{s^2 - \beta^2}
  ds.
\end{equation}
Equation (\ref{d_j_step1}) produces a recursion relation for
$d_j$:
\begin{equation} \label{d_j_step2}
d_j = \frac{1}{2} \frac{1}{j+1} \lambda^{2j-1} (\lambda^2 -
\beta^2) \sqrt{\lambda^2 - \beta^2} + \frac{1}{2}
\frac{(2j-1)}{j+1} \beta^2 d_{j-1}.
\end{equation}

The initial condition of this recursion is
\begin{equation} \label{d_0}
d_0 = \int_\beta^\lambda \sqrt{\lambda^2 - \beta^2} d\lambda =
\frac{1}{2} \lambda \sqrt{\lambda^2 - \beta^2} - \frac{1}{2}
\beta^2 \log\left| \frac{\lambda}{\beta} + \sqrt{\lambda^2 -
\beta^2} \right|.
\end{equation}

The recursion relation (\ref{d_j_step2}), (\ref{d_0}) is solved by
\begin{eqnarray} \index{$S^{(i)}_j(\lambda)$}
d_j &=& S^{(1)}_j(\lambda) (\lambda^2 - \beta^2) \frac{ \sqrt{
\lambda^2 -
    \beta^2}}{\lambda} + S^{(2)}_j(\lambda) \frac{ \sqrt{ \lambda^2 -
      \beta^2 }}{\lambda}
\\ \nonumber & &
       + S^{(3)}_j(\lambda) \log\left(
    \frac{\lambda}{\beta} + \frac{\sqrt{\lambda^2 -
    \beta^2}}{\beta}\right), \, j \in \mathbb{N}.
\end{eqnarray}

The functions $S_j^{(2)}$ and $S_j^{(3)}$ are simple. We find the
expressions:
\begin{eqnarray}
S_j^{(2)} &=& \frac{ \beta^{2j} }{ 4^j } \binom{2j -1}{j-1}
\frac{1}{j+1} \lambda^2
= \frac{ v_j}{\beta^2} \lambda^2, \\
S_j^{(3)} &=& -\frac{ \beta^{2j} }{ 4^j } \binom{2j -1}{j-1}
\frac{1}{j+1} = -v_j.
\end{eqnarray}

The case of $S^{(1)}_j$ is more complicated;  after some work we have
\begin{equation}
S^{(1)}_j = \frac{ 1 }{ 2^j }\binom{2j-1}{j-1} \frac{1}{j+1}
\sum_{i=1}^j \frac{2^{j-2i+1} (j-i+2) \beta^{2i-2} }{ \binom{2j
-2i +1}{j-i} } \lambda^{2j -2i + 2} = \frac{  v_j }{2}
\sum_{i=1}^j \frac{ \lambda^{2i}}{ v_i (i+1) } .\end{equation} The
integral (\ref{4_25}), in terms of $S_j^{(1)}$, $S_j^{(2)}$, and
$S_j^{(3)}$ is
\begin{eqnarray}
\int_\beta^\lambda h(s) \sqrt{ s^2 - \beta^2} ds &=& \frac{1}{x}
\left( d_0 + \sum_{j=0}^{\nu-1} h_j d_j \right)
\\ &=&
\frac{1}{x} \left( \sum_{j=0}^{\nu-1} h_j S_j^{(1)} \right) \left( \lambda^2 -
\beta^2 \right) \frac{ \sqrt{ \lambda^2 - \beta^2}}{\lambda}
\\ \nonumber & &
+ \frac{1}{x} \left( S_0^{(2)}
  + \sum_{j=0}^{\nu-1} h_j S_j^{(2)} \right) \frac{ \sqrt{ \lambda^2 -
      \beta^2}}{\lambda} +  \\  & & \label{4_31}
\frac{1}{x} \left( S_0^{(3)} + \sum_{j=0}^{\nu-1} h_j S_j^{(3)} \right)
\log\left( \frac{\lambda}{\beta} + \frac{\sqrt{\lambda^2 -
\beta^2}}{\beta}\right).
\end{eqnarray}
Using the relation (\ref{con1}) and the explicit expressions for
the $S_j^{(k)}$ derived above, we finally arrive at the form of
the expansion for (\ref{4_25}) given in (\ref{intexp}).

\bibliographystyle{amsplain}

\end{document}